\documentclass{article}
 
\usepackage{makeidx}  
\usepackage{graphicx} 
\usepackage{amsmath}
\usepackage{amssymb}

\def\eq#1{Eq.~(\ref{#1})}
\def\fig#1{Fig.~\ref{#1}} 
\def\tab#1{Tab.~\ref{#1}}
\def\sec#1{Sec.~\ref{#1}}  
 
\def\G{{\cal G}}
\def\H{{\cal H}}   
\def\x{\textbf{x}}
\def\y{\textbf{y}}
\def\z{\textbf{z}}
\def\t{\textbf{t}}           
\def\r{\textbf{r}}

\begin{document}   

\title{Bicontinuous surfaces in self-assembling amphiphilic systems}

\author{U. S. Schwarz$^{1}$ and G. Gompper$^{2}$\\
{\small ${}^1$Max-Planck-Institut f\"ur Kolloid- und Grenzfl\"achenforschung, D-14424 Potsdam}\\
{\small ${}^2$Institut f\"ur Festk\"orperforschung, Forschungszentrum J\"ulich, D-52425 J\"ulich}}
\date{ }

\maketitle

\begin{center}
\fbox{\parbox{0.9\textwidth}{Appeared in \textit{Morphology of Condensed 
Matter: Physics and Geometry of Spatially Complex Systems},
eds. K. R. Mecke and D. Stoyan, Springer Lecture Notes in Physics
vol. \textbf{600}, pp. 107-151 (2002)}}
\end{center}                                                        

\begin{abstract}
  Amphiphiles are molecules which have both hydrophilic and hydrophobic parts.
  In water- and/or oil-like solvent, they self-assemble into extended
  sheet-like structures due to the hydrophobic effect. The free energy
  of an amphiphilic system can be written as a functional of its
  interfacial geometry, and phase diagrams can be calculated by
  comparing the free energies following from different geometries.
  Here we focus on bicontinuous structures, where one highly convoluted
  interface spans the whole sample and thereby divides it into two
  separate labyrinths. The main models for surfaces of this class are
  triply periodic minimal surfaces, their constant mean curvature and
  parallel surface companions, and random surfaces. We discuss the
  geometrical properties of each of these types of surfaces and how
  they translate into the experimentally observed phase behavior of
  amphiphilic systems.
\end{abstract}  

\section{Surfaces in self-assembling amphiphilic systems}
\label{sec:intro}

The subject of this article is the theoretical description of
amphiphilic systems, which are one example of \emph{soft condensed
  matter}.  Soft matter is material that has a typical energy scale of
$k_B T \approx 4\ \times 10^{-21} J$, where $k_B$ is the Boltzmann
constant and $T \approx 300 K$ is room temperature. For such material,
perturbations arising from the thermally activated movement of its
molecular components are sufficient to induce configurational changes,
and entropy is at least equally important as energy in determining its
material properties.  The understanding of the underlying mechanisms
is essential for the application of many technologies in everyday
life, including colloidal dispersions (paints, inks, food, creams,
lotions), foams (beverages), liquid crystals (displays),
polyelectrolyte gels (diapers) and soaps (washing and cleaning).
Moreover, the term \emph{soft matter} also includes most biomaterials, for
example blood or cartilage.  One particularly important biomaterial is
the \emph{biomembrane}, that is the protein carrying lipid bilayer
which surrounds each cell and its organelles. A good model system for
the structural properties of biomembranes are \emph{lipid bilayers},
which form spontaneously in mixtures of water and lipids and which are
one of the main subjects of this paper.

Soft matter systems are very often characterized by competing
interactions (including entropic ones), which lead to structure
formation on the length scale between tens and hundreds of nanometers
(1\ nm = $10^{-9}$ m). Since this length scale is not accessible by
optical lithography, self-assembly in soft matter systems is one of
the main concepts of nanoscience. With structure
formation being so prominent in soft matter systems, their theoretical
description often centers around their spatial structure in three
dimensions. For a rough classification of the different approaches
used, it is convenient to use an analogy to the field of random
geometries, and to distinguish between point, fiber and surface processes.
Examples of systems which are suited for approaches along these lines
are colloidal dispersions, polymer networks and sheet-like structures
in self-assembling amphiphilic systems, respectively. Here we will
treat the latter case, but we will also show that surface dominated
system can be related to Gibbs distributions for scalar fields,
namely through the use of isosurface constructions.

\begin{figure}
\begin{center}
\includegraphics[width=\textwidth]{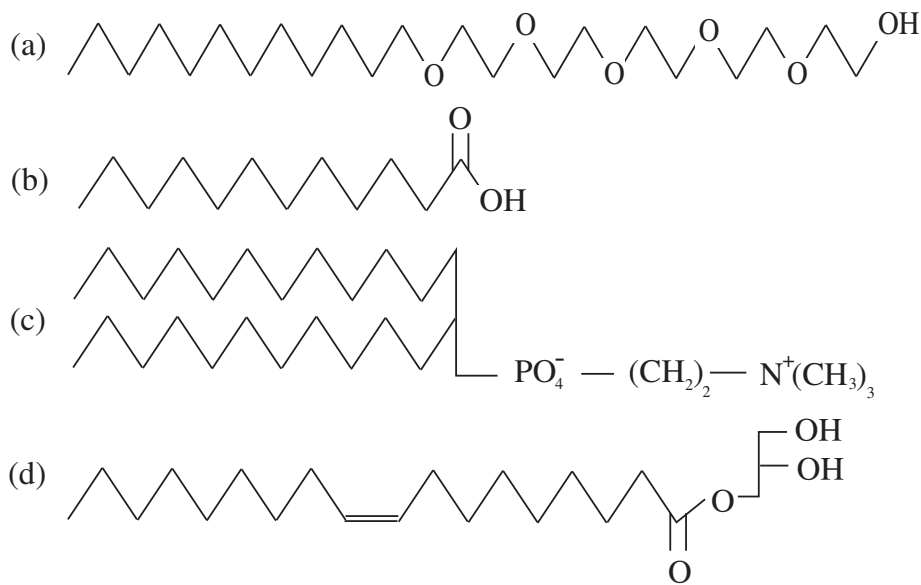}
\caption{\label{fig:amphiphiles}Schematic representation of different amphiphilic
molecules. Hydrophobic tails are to the left, hydrophilic heads to the
right. (a) Pentaethylene glycol dodecyl ether,
$C_{12}H_{25}(OCH_2CH_2)_5OH$ or in short $C_{12}E_{5}$, is a small
surfactant. (b) Lauric acid, $CH_3(CH_2)_{10}COOH$ or in short LA.
Fatty acids are the simplest amphiphiles and can be dissolved to high
amounts in phospholipid bilayers.  (c) Dilauroyl phosphatidylcholine,
or in short DLPC.  Phosphatidylcholines have two hydrocarbon tails and
a zwitterionic headgroup.  They are abundant in animal cells and the
best studied model system for biological lipids. (d) Monoolein, a
monoacyl glycerol with one (unsaturated) hydrocarbon tail.}
\end{center}
\end{figure}       
       
\emph{Amphiphilic systems} are solutions of amphiphiles in suitable
solvent. \emph{Amphiphiles} are molecules which consist of two parts,
one being hydrophilic (water-like, called the head) and one being
hydrophobic (oil-like, called the tail). Well known classes of
amphiphiles are \emph{tensides} (used for washing and cleaning
purposes) and \emph{lipids} (the basic components of biomembranes).
\fig{fig:amphiphiles} shows several examples for which phase
diagrams are discussed below.  Tensides are
often called \emph{surfactants}, due to their \underline{surf}ace
\underline{act}ivity at interfaces between water-like and oil-like phases.
Hydrophilic and hydrophobic molecules demix due to the hydrophobic
effect: oil-like molecules are expelled from a region of water-like
molecules since they disturb their network of hydrogen
bonds. Hydrophilic and hydrophobic parts of an amphiphilic molecule
cannot demix due to the covalent linkage between them, but the
amphiphiles can self-assemble in such a way as to shield its
hydrophilic parts from its hydrophobic ones and vice
versa. Amphiphilic systems have many similarities with diblock
copolymer systems, for which the term \emph{microphase separation} is
used to denote the molecular tendency for structure formation (see the
contribution by Robert Magerle). However, in contrast to diblock
copolymers, amphiphiles are small molecules, thus the entropy of their
molecular configurations plays only a minor role in determining the
overall structure. Moreover, whereas in diblock copolymer systems one
component is sufficient to obtain stable mesophases, in amphiphilic
systems usually the presence of an aqueous solvent is necessary to
obtain well-pronounced structure formation. Both diblock copolymer and
amphiphilic systems show structure formation on the nanometer scale,
but the length scale is set by different control parameters: for
diblock copolymers and amphiphiles, these are polymer size and solvent
concentrations, respectively.  Since the solvent in an amphiphilic
system usually has no special physical properties by itself, its
properties are mostly determined by its interfaces. This stands in
marked contrast to the case of diblock copolymer systems, where the
entropy of chain configurations in the regions away from the
interfaces cannot be neglected in a physical description (for a review
see \cite{a:mats96} and references therein). As we will discuss in
more detail below, interface descriptions have been very successful in
describing the properties of amphiphilic systems (for reviews see
\cite{a:lipo91,a:Gomp94,a:lipo95c,a:seif96a,a:safr99}).

\begin{figure}
\begin{center}
\includegraphics{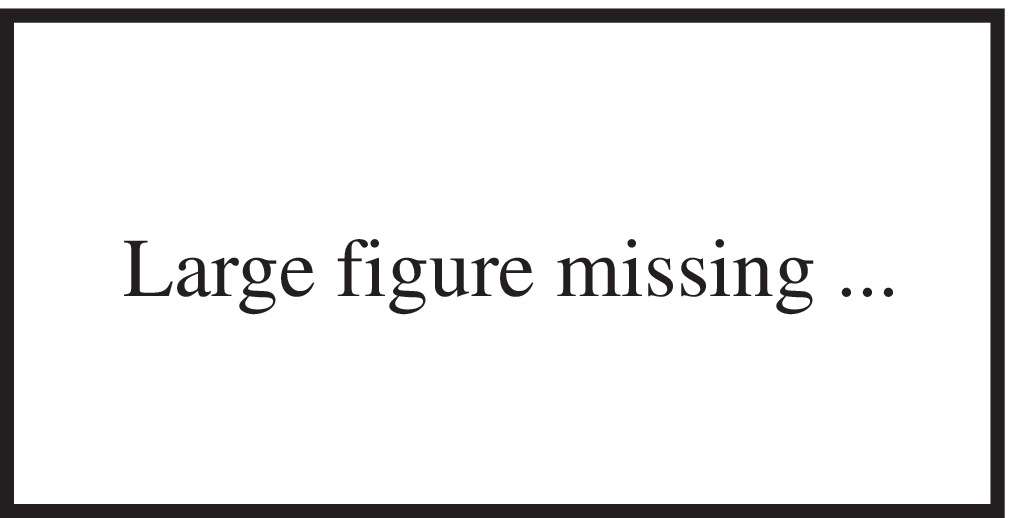}
\caption{\label{fig:selfassembly}Self-assembly of amphiphiles: 
(a) in binary systems, amphiphiles self-assemble into bilayers,
micelles and vesicles in order to shield the hydrophobic tails from
the aqueous solvent. (b) In ternary systems, the oil-like solvent
swells the hydrophobic regions. If the oil-like solvent is the
majority component, inverse structures occur.  In ternary systems, all
interfaces are amphiphilic monolayers.}
\end{center}
\end{figure}

Amphiphilic systems can be classified according to the solvent used. A
mixture of water and amphiphile is called a \emph{binary} system, and
a mixture of water, oil and amphiphile is called a \emph{ternary}
system.  In binary systems, amphiphiles self-assemble into spherical,
cylindrical or bilayer structure in such a way that the hydrophobic
tails are shielded from the hydrophilic solvent; in ternary systems,
amphiphiles self-assemble into monolayer structures in such a way that
the hydrophobic tails and the hydrophilic heads face the hydrophobic
and hydrophilic solvents, respectively (see
\fig{fig:selfassembly}). Irrespective of these differences,
amphiphiles in binary and ternary systems assemble into similar
geometries, because in both cases one deals with extended sheet-like
structures (the same geometries also occur in diblock copolymer
systems).  At room temperature, amphiphiles in mono- and bilayers
usually form a two-dimensional fluid, that is the molecular
constituents show only short-ranged and no long-ranged order in the
plane of the interface.  Therefore the sheet-like structures in
amphiphilic systems are called \emph{fluid membranes} (in fact
membrane fluidity is essential for the functioning of the protein
machinery carried by biomembranes).

\begin{figure}
\begin{center}
\includegraphics{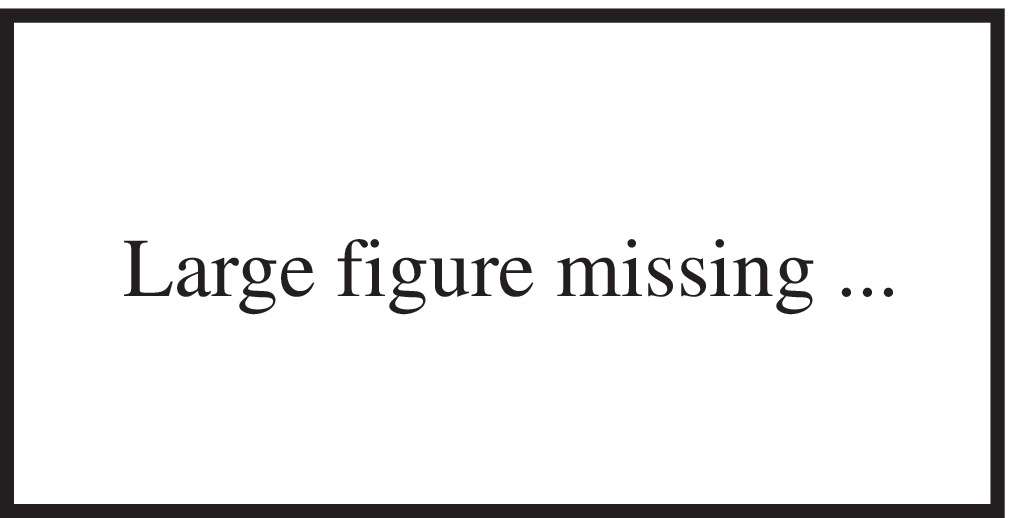}
\end{center}
\caption{\label{fig:geometries}Geometries of self-assembled interfaces: (a) 
micellar (spheres), (b) hexagonal (cylinders), (c) lamellar (planes),
(d) cubic bicontinuous (cubic minimal surfaces) and (e) disordered
bicontinuous (random surfaces). In amphiphilic systems,
mono- or bilayers are draped onto the mathematical surfaces.
\emph{Bicontinuous} means that there is one surface which spans 
the whole sample, thereby separating it into two disjunct yet
intertwined labyrinths (like the corresponding surface, each labyrinth
is connected and spans the whole sample). Note that except for (e),
all surfaces are ordered and have constant mean curvature.}
\end{figure} 

Since the physics of amphiphilic systems is mostly determined by their
interfaces, different geometries have comparable free energies and
small changes in external variables can induce phase transitions.  In
particularly, for amphiphilic systems different geometries are stable
for different values of concentrations and temperature. In order to
achieve systematic control of amphiphilic systems, a large effort has
been invested into experimentally determining the phase diagram for many 
important amphiphilic systems. It has been found that despite their
molecular diversity, the phase behavior of amphiphilic systems follows
some general rules which result from their intrinsic tendency for
self-assembly into sheet-like structures. For example, for increasing
amphiphile concentration one usually finds the generic phase sequence
micellar disordered - micellar ordered - hexagonal - cubic
bicontinuous - lamellar. \emph{Micellar} means spherical geometry;
\emph{micellar ordered} corresponds to an ordered (usually cubic)
array of spherical micelles. \emph{Hexagonal} denotes a
two-dimensional packing of cylindrical aggregates. \emph{Bicontinuous}
means that one surface partitions space into two separate labyrinths,
each of which can be used to traverse space. \emph{Cubic bicontinuous}
corresponds to a space-filling arrangement of one interface, which
folds onto itself in a cubic arrangement. Sometimes non-cubic
spacegroups are found, but most ordered bicontinuous structures in
equilibrium are cubic. In some systems, bicontinuous yet disordered
phases occur, which are called \emph{sponge phases} and
\emph{microemulsions} for binary and ternary systems, respectively.
\emph{Lamellar} corresponds to a one-dimensional stack of interfaces.
Ordered arrangements like hexagonal or cubic symmetries can be probed
by scattering techniques, and bicontinuity by diffusion experiments
(e.g.\ using nuclear magnetic resonance).  \fig{fig:geometries} tries
to visualize the different interface geometries.  It is the case
of cubic bicontinuous phases which is the main focus of this paper.

\begin{figure}
\begin{center}
\includegraphics{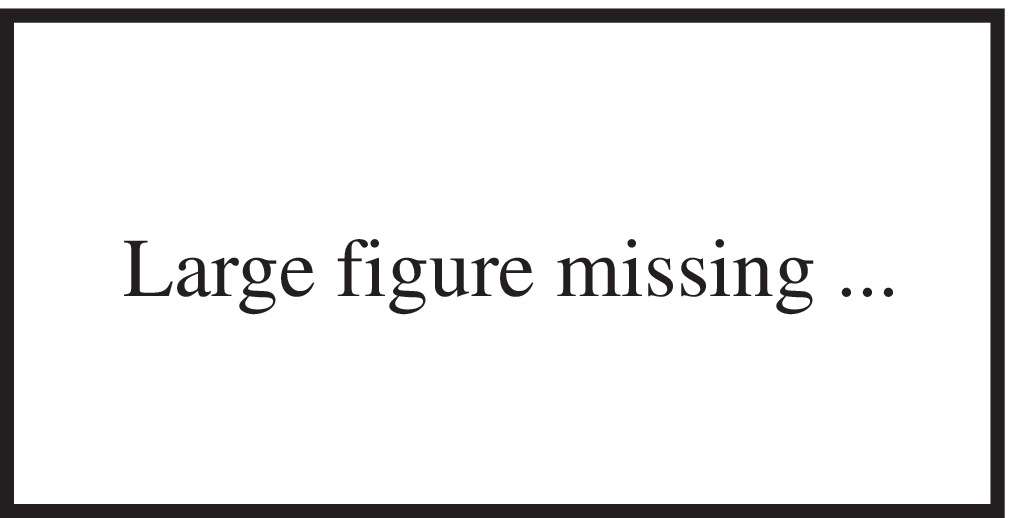}
\end{center}
\caption{\label{fig:phasebehavior1}Experimentally determined phase behavior for the
surfactant $C_{12}E_{5}$: binary system ($H_20 / C_{12}E_{5}$) 
\cite{a:stre90} (top) and ternary system ($H_20 / C_{12}E_{5} / C_{14}$)
\cite{a:olss93} (bottom). In the binary system, the hexagonal phase $H_1$,
the cubic bicontinuous phase $V_1$, the lamellar phase $L_{\alpha}$,
the sponge phase $L_3$ and the micellar phases $L_1$ and $L_2$ are
stable. In the ternary system, the same phases occur close to the
binary side $H_20 / C_{12}E_5$.  At $T_1 = 5^{\circ} T$ and $T_2 =
25^{\circ} T$, $L_1$ coexists with excess oil (emulsification
failure). At $T_3 = 48^{\circ} T$ (balanced temperature), a
bicontinuous microemulsion is stable in the middle of the Gibbs
triangle.}
\end{figure}

There exists a large body of literature on the occurrence of
bicontinuous phases in amphiphilic systems (for reviews see
\cite{a:font90,a:luzz93,a:sedd95}).  In \fig{fig:phasebehavior1} we
show experimental phase diagrams for binary \cite{a:mitc83,a:stre90}
and ternary \cite{a:kuni82,a:olss93} systems with the small surfactant
$C_{12}E_{5}$ (this is the molecule shown in \fig{fig:amphiphiles}a).  
For the binary system, \fig{fig:phasebehavior1}a, at
low temperature one sees the sequence hexagonal $H_1$ - cubic
bicontinuous $V_1$ - lamellar $L_{\alpha}$ with increasing amphiphile
concentration. For high temperature, the hexagonal and cubic
bicontinuous phases disappear, the lamellar phase $L_{\alpha}$ expands
(its interfaces unbind), and for very small amphiphile concentration a
sponge phase $L_3$ occurs. For the ternary system,
\fig{fig:phasebehavior1}b, we recognize the same situation again close
to the binary side water-amphiphile. For equal amounts of water and
oil and not too large concentration of amphiphile, that is in the
lower middle of the Gibbs triangle, one sees the following phase
behavior: at low temperature an emulsification failure occurs, that is
the micellar disordered phase $L_1$ coexists with an oily excess
phase. At high temperature, the emulsification failure disappears and
is replaced by a microemulsion (the central, triangular one phase
region). We can conclude that at low temperature, strongly curved
structures prevail ($H_1$, $L_1$), while at high temperature, those
structures become stable which are locally flat ($L_{\alpha}$, $L_3$,
microemulsion).  As we will show below, this behavior can be
understood using the concept of a temperature dependent spontaneous
curvature: spontaneous curvature is finite at low temperature, but
vanishes at high temperature.  The concept of finite spontaneous
curvature also offers a natural explanation for the emulsification
failure at low temperature.  As we will argue below, the cubic
bicontinuous phase $V_1$ is an intermediate structure between strongly
curved and locally flat. It also follows from the phase behavior
described that at low temperature, stable phases tend to be ordered
($H_1$, $V_1$), while disordered phases profit from increased
temperature ($L_3$, microemulsion). We will discuss below that the
disordered phases indeed have a larger configurational entropy. In
this respect, it is the lamellar phase $L_{\alpha}$ which acts as an
intermediate structure.

\begin{figure}
\begin{center}
\includegraphics{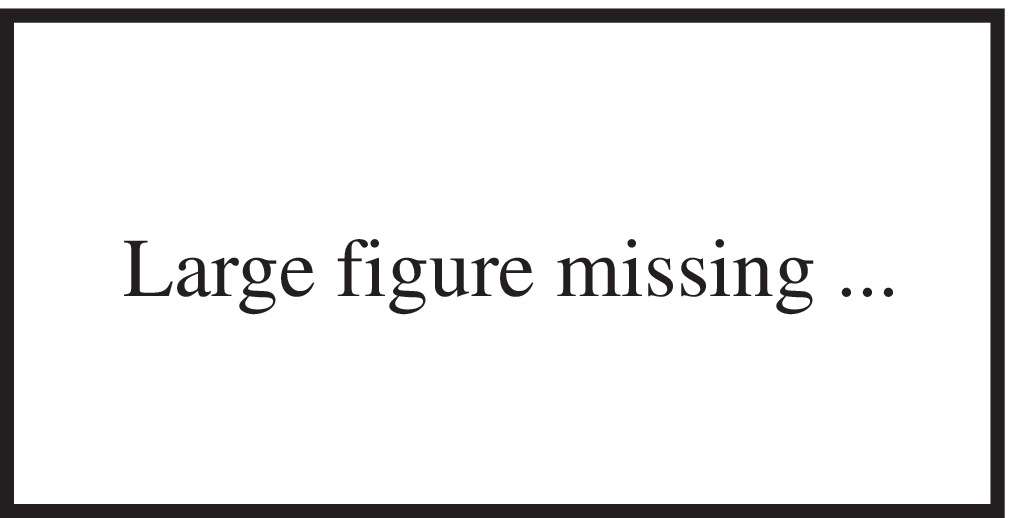}
\end{center}
\caption{\label{fig:phasebehavior2}Experimentally determined phase behavior for lipid-water
mixtures: 2:1 lauric acid / dilauroyl phosphatidylcholine
\cite{a:temp98a} (top) and monoolein \cite{a:qiu00} (bottom). In both phase
diagrams, one sees the sequence lamellar - cubic bicontinuous - excess
water (emulsification failure) at intermediate temperatures (around
$40^{\circ} C$) and with increasing water concentration.  In (a), G, D
and P are stable. In (b), G (Ia3d) and D (Pn3m) are stable.}
\end{figure}

In \fig{fig:phasebehavior2} we show experimental phase diagrams for
two different lipid-water mixtures. \fig{fig:phasebehavior2}a is for water and 2:1
lauric acid / dilauroyl phosphatidylcholine \cite{a:temp98a}
and \fig{fig:phasebehavior2}b for water and monoolein
\cite{a:qiu00} (these are the molecules shown in \fig{fig:amphiphiles}b-d). 
Despite the molecular differences, the macroscopic phase behavior is
surprisingly similar.  At low temperature the membranes loose their
fluidity. At intermediate temperatures, the lamellar phase is stable,
and at high temperatures, it is replaced by the hexagonal
phase. Whereas in the case of the surfactant $C_{12}E_{5}$ spontaneous
curvature is finite at ambient temperatures and vanishes at high
temperature, for the lipids spontaneous curvature vanishes at ambient
temperatures and increases with temperature. Several cubic
bicontinuous phases are stable for intermediate temperatures, in the
sequence lamellar - G - D - P (here G, D and P stand for cubic
bicontinuous structures which are discussed in more detail below) with
increasing water content. The last phase in this sequence undergoes an
emulsification failure, that is it coexists with an excess water
phase. We will show below that this phase behavior can be explained
nicely if the lipid monolayers of the cubic bicontinuous phases are
modeled as parallel surfaces to a cubic minimal midsurface.

In this article, we focus on the interface description of amphiphilic
systems, that is the free energy of the system is written as a
functional of its interface configuration. The interface free energy
is introduced in the next section (\sec{sec:curvature}), and in the
rest of this article we will specify this free energy expression for
different instances of cubic bicontinuous phases. In each case, we
will discuss the relevant geometric properties and show how they
relate to the resulting free energy expressions and phase diagrams.
Our starting point are \emph{triply periodic minimal surfaces} (TPMS)
in \sec{sec:TPMS}, a subject well-known from differential
geometry. These structures are expected to occur if the amphiphilic
interface is symmetric in regard to its two sides and if temperature
is not too high as to destroy the ordered state. TPMS are also the
reference state for their parallel surfaces and CMC-companions, which
are the adequate structural models for a detailed analysis of
amphiphilic monolayers in cubic bicontinuous phases.  In the case that
the physical interface is not symmetric in regard to its two sides
(that is if spontaneous curvature exists, like usually for an
amphiphilic monolayer at a water-oil interface), the relevant
mathematical representations are the constant mean curvature
companions of the TPMS, that is \emph{triply periodic surfaces of
constant mean curvature} (CMC-surfaces) treated in
\sec{sec:cmc}.  The case of lipid bilayers might be considered to be
the composition of two such CMC-surfaces, but we will argue in
\sec{sec:parallel} that in this case the relevant geometry is in fact
the one of \emph{parallel surfaces} to a TPMS. In all of these cases,
we are interested in ordered structures, and therefore the main method
will be minimization of the corresponding free energy
functionals. This is different in \sec{sec:random}, where we will
discuss disordered bicontinuous phases, that is the sponge phases and
microemulsions, which often occur at higher temperatures due to
entropic effects. These structures are modeled as \emph{random
surfaces}, thus the main model here will be Gibbs distributions, in
particular the theory of Gaussian random fields and Monte Carlo
simulations.

\section{Free energy functionals}
\label{sec:curvature}

\subsection{Interface models}
\label{sec:helfrich}

From the viewpoint of elasticity theory, amphiphilic interfaces with
small curvatures can be considered to be thin elastic shells, which
are known to have few fundamental modes of deformation: out-of-plane
bending, in-plane compression and in-plane shearing \cite{b:land70}.
However, since amphiphilic interfaces are fluid and nearly
incompressible, in-plane strain is irrelevant and the most relevant
deformation mode is bending. For small curvatures the free energy of
an amphiphilic interface is a function only of its geometry
\cite{a:canh70,a:helf73}:
\begin{equation} \label{eq:Helfrich}
F = \int dA \left\{ \sigma + 2 \kappa (H - c_0)^2 + \bar \kappa K \right\}\ .
\end{equation} 
Here $dA$ denotes the differential area element and $H$ and $K$ mean
and Gaussian curvature, respectively.  The latter two follow from the
two principal curvatures $k_1$ and $k_2$ as $H = (k_1 + k_2) / 2$ and
$K = k_1 k_2$.  The three material parameters introduced in
\eq{eq:Helfrich} define the energy scales of the corresponding changes:
$\sigma$ is \emph{surface tension} and corresponds to changes in
surface area, $\kappa$ is \emph{bending rigidity} and corresponds to
cylindrical bending, and $\bar \kappa$ is \emph{saddle-splay modulus}
(or \emph{Gaussian bending rigidity}) and corresponds to changes in
topology due to the Gauss-Bonnet-theorem for closed surfaces, $\int dA
K = 2 \pi \chi$, where $\chi$ denotes Euler characteristic. The
\emph{spontaneous curvature} defines the reference point for bending
deformations.  It has to vanish for symmetric amphiphilic sheets (like
lipid bilayers), but in general is finite for non-symmetric ones (like
monolayers). For monolayers, it is usually a linear function of
temperature, $c_0 \sim (T - T_b)$. Therefore spontaneous curvature
vanishes at the balanced temperature $T_b$, at which solvent
properties make the monolayer symmetric in regard to bending.

For amphiphilic interfaces, surface area is proportional to the number
of amphiphilic molecules, thus $\sigma$ can also be interpreted as a
chemical potential for amphiphiles.  The bending rigidity $\kappa$ has
to be positive, otherwise the system would become instable to
spontaneous convolutions. For surfactant and lipid systems, its values
are of the order of 1 and 20 $k_B T$, respectively. Surface tension
$\sigma$ can assume negative values (favorable chemical potential for
the influx of amphiphiles), as long as bending rigidity $\kappa$
exists and has a positive value, in order to prevent an instability
towards proliferation of interfacial area. The value of the
saddle-splay modulus $\bar \kappa$ is often debated, but usually it is
assumed to have a small negative value. This assumption is validated
by the following argument \cite{a:helf81}: for $\sigma = 0$ and $c_0 =
0$, we can rewrite \eq{eq:Helfrich} as
\begin{equation} \label{Helfrich2}
F = \int dA \left\{ \frac{1}{2} \kappa_+ (k_1 + k_2)^2 
  + \frac{1}{2}\kappa_- (k_1 - k_2)^2 \right\}
\end{equation}
with
\begin{equation}
\kappa_+ = \kappa + \frac{\bar \kappa}{2}, \quad
\kappa_- =  - \frac{\bar \kappa}{2} \ .
\end{equation}
From this we conclude that the Gaussian bending rigidity has to
satisfy $\kappa_+ > 0$ and $\kappa_- > 0$, that is $- 2 \kappa < \bar
\kappa < 0$, otherwise the system would become unstable. For $\bar
\kappa < - 2 \kappa$ ($\kappa_+ < 0$), we would get $k_1 = k_2 \to
\infty$, that is many small droplets, and for $\bar \kappa > 0$ 
($\kappa_- < 0$), we would get $k_1 = - k_2 \to \infty$, that is a
minimal surface with very small lattice constant. 

From the mathematical point of view, it is interesting that for
$\sigma = 0$ and $c_0 = 0$, the bending energy from \eq{eq:Helfrich}
is not only invariant under rescaling with length, but also invariant
under conformal transformations in general. This has intriguing
consequences for vesicles \cite{a:seif96a} (vesicles are depicted in
\fig{fig:selfassembly}, but are not subject of this article) and phase
transitions in systems without spontaneous curvature (as will be
discussed in \sec{sec:random}).

Since the energy scales involved are of the order of $k_B T$, thermal
noise is sufficient to induce shape changes. Due to thermal
fluctuations on smaller length scales, the effective values for the
material parameters are changed (renormalized) at larger length
scales \cite{helf85,peli85}.  It has been shown in 
the framework of renormalization group
theory that logarithmic corrections arise due the two-dimensional
nature of the amphiphilic interfaces, so that 
\cite{peli85,davi89a,cai94}
\begin{align} \label{eq:renorm_kappa}
\kappa_R(l) & = \kappa - \frac{3 k_B T}{4 \pi} \ln \frac{l}{\delta}\ , \\
\label{eq:renorm_kbar}
\bar\kappa_R(l) & = \bar \kappa 
            + \frac{5 k_B T}{6 \pi} \ln \frac{l}{\delta}\ 
\end{align}
where $l$ is the length scale on which the system is analyzed and
$\delta$ is a microscopic cutoff on the scale of the membrane
thickness.  Thus bending rigidity decreases, while Gaussian bending
rigidity increases with increasing length scale $l$.

For the following, it is useful to introduce two dimensionless
quantities which correspond to the two curvatures of a two-dimensional
surface. For a structure with surface area $A$ and volume $V$, the
length scale $V / A$ can be used to rescale its curvatures. In order
to be able to convert easily from local to global quantities, we
consider a triply periodic CMC-surface with Euler characteristic
$\chi$ and integrated mean curvature $H_i = \int dA H$ per
conventional (that is simple cubic) unit cell:
\begin{equation} \label{eq:dimlesscurv}
K \left( \frac{V}{A} \right)^2 = \frac{2 \pi \chi V^2}{A^3}\ , \quad 
H \left( \frac{V}{A} \right) = \frac{H_i V}{A^2}\ . 
\end{equation}
These expressions motivate the definition of the \emph{topology index} 
$\Gamma$ and the \emph{curvature index} $\Lambda$:
\begin{equation} \label{eq:indices}
\Gamma = \left( \frac{{A^*}^3}{2 \pi |\chi|} \right)^\frac{1}{2}\ , \quad
\Lambda = \frac{H^*_i}{{A^*}^2}
\end{equation}
where $A^* = A / V^{2/3}$ and $H^*_i = H_i / V^{1/3}$ are the scaled
surface area and the scaled integral mean curvature per conventional
unit cell, respectively.  Note that these definitions can be applied
to any triply periodic surface; here we used CMC-surfaces only for an
heuristic motivation of the quantities defined. Both the topology
index $\Gamma$ and the curvature index $\Lambda$ do not depend on
scaling and choice of unit cell. They are universal geometrical
quantities which characterize a surface in three-dimensional space.
The topology index describes its porosity (the larger its value, the
less holes the structure has) and its specific area content (the
larger its value, the more inner surface the structure contains), and
the curvature index describes how strongly the structure is curved
(irrespective of the actual lattice constant). It is interesting to
note that the two indices defined here correspond to the two
isoperimetric ratios known from integral geometry. For a TPMS,
the curvature index $\Lambda$ vanishes and the topology index 
$\Gamma$ is its most important geometric characteristic.

\subsection{Ginzburg-Landau models}
\label{sec:Landau}

A different but equally powerful approach to amphiphilic interfaces is
the isosurface construction (also known as phase field method), which
derives a two-dimensional surface from a three-dimensional scalar
field $\Phi(\r)$. For ternary amphiphilic systems, $\Phi(\r)$ can be
interpreted as the local concentration difference between water ($\Phi
= 1$) and oil ($\Phi = -1$). The position of the amphiphilic monolayer
can be identified with the isosurface $\Phi(\r) = 0$. For binary
amphiphilic systems, $\Phi(\r)$ can be interpreted as the local
concentration difference between water on different sides of a
bilayer, and the isosurface $\Phi(\r) = 0$ marks the position of the
bilayer mid-surface. In both cases, an energy functional can be
defined in the spirit of a Ginzburg-Landau theory for the scalar
field $\Phi(\r)$. A reasonable choice is the $\Phi^6$-model
introduced by Gompper and Schick \cite{a:gomp90,a:Gomp94}:
\begin{equation} \label{Phi6}
{\cal F}[\Phi] = \int\ d\textbf{r}\ \left\{ (\Delta \Phi)^{2}
+ g(\Phi) (\nabla \Phi)^{2} + f(\Phi) \right\}\ 
\end{equation}
with the following choice for $f$ and $g$:
\begin{equation} \label{GL_density}
f(\Phi) = (\Phi + 1)^{2} (\Phi - 1)^{2} (\Phi^{2} + f_0)\ ,\                  
g(\Phi) = g_0 + g_2 \Phi^{2}\ .
\end{equation}
It has been shown that this model is similar to an interface
description as given in \eq{eq:Helfrich}, and prescriptions have been
given how to calculate the parameters of the interface
Hamiltonian from the given Ginzburg-Landau theory \cite{a:gomp92a}.
Due to the invariance under $\Phi \to - \Phi$, spontaneous curvature
$c_0$ vanishes in this model.  For given model parameters $(g_0, g_2,
f_0)$, this energy functional can be minimized for its spatial degrees
of freedom, and a phase diagram can be calculated as a function of
model parameters by identifying the absolute minimum at every point. A
reasonable choice for amphiphilic systems is $g_0 < 0$, $f_0$ close to
$0$ and $g_2$ not too large. Then the formation of interfaces is
favored and a lamellar phase becomes stable. Calculation of the
corresponding parameters of the interface model then yields $\sigma <
0$, $\kappa > 0$ and $\bar \kappa < 0$. Note that $\sigma < 0$ favors
the formation of interfaces, and that $\bar \kappa < 0$ favors the
lamellar phase.

\section{Triply periodic minimal surfaces (TPMS)}
\label{sec:TPMS}

Minimal surfaces and surfaces of constant mean curvature (see
\sec{sec:cmc}) in general have attracted a lot of attention both 
in mathematics and in physics, partially due to their intrinsic
beauty, which might be considered to follow from the fact that they
are solutions to variational problems \cite{a:hild96}.  It is
important to note that the different fields are interested in
different aspects of the same object: while for mathematicians the
most central question is the existence proof for a minimal surface of
interest, physicists are more interested in its representation, which
can be used to derive physical properties of corresponding material
systems. As Karcher and Polthier remark, outside mathematics only
pictured minimal surfaces have been accepted as existent
\cite{a:karc96}. In fact this statement can also be reversed: not
every structure which one can picture is necessarily a minimal
surface. In this article we consider only established minimal
surfaces. As we are motivated by physical considerations, we are
interested only in embedded surfaces which do not self-intersect. In
regard to mathematics, our main emphasis here will be on
representations, since the physical properties of amphiphilic systems
follow from their spatial structure.

Minimal surfaces are surfaces with $H = 0$ everywhere. If one varies a
surface with a normal displacement $\delta \phi(u,v)$ (where $u$ and
$v$ are the internal coordinates of the surface), it can be shown that
the change in surface area is $\Delta A = 2 \delta \int dA \phi(u,v)
H(u,v) + O(\delta^2)$ \cite{a:jost94}, that is a minimal surface is a
stationary surface for variations of surface area. Therefore surfaces
under surface tension (like soap films), which try to minimize surface
area, form minimal surfaces. Here we consider surfaces which are
dominated by bending rigidity rather than by surface tension. However,
for vanishing spontaneous curvature the resulting structures also
correspond to minimal surfaces. \eq{eq:Helfrich} explains why: 
these surfaces have to minimize $\int dA
H^2$ (they are so-called \emph{Willmore surfaces}), and minimal
surfaces are a special case of Willmore surfaces, since $H = 0$ is a
trivial minimization of this functional.

If mean curvature $H = (k_1 + k_2) / 2 = 0$, then Gaussian curvature
$K = k_1 k_2 = - k_1^2 \leq 0$.  Thus minimal surfaces are everywhere
either saddle-like or flat, and nowhere convex. In fact it can be
shown that the flat points with $k_1 = k_2 = 0$ are isolated. Moreover
it follows from $K \leq 0$ that minimal surfaces without boundaries
cannot be compact \cite{a:jost94}.  Until 1983, only two embedded
minimal surfaces were known which are non-periodic: the plane and the
catenoid.  Then a new surface of this type was found, the
Costa-surfaces, which from the distance looks like the union of a
plane and a catenoid \cite{a:karc96}.  Today some more surfaces of
this type are known, but the majority of all embedded minimal surfaces
without boundaries are in fact periodic.  There is only one
simply-periodic minimal surface, the helicoid, and there are some
doubly-periodic surfaces, with the Scherk-surface being most
prominent. The majority of all known periodic minimal surfaces is
triply-periodic. Here we will focus on cubic minimal surfaces and
their physical realizations, which are cubic bicontinuous structures.

\begin{figure}
\begin{center}
\includegraphics{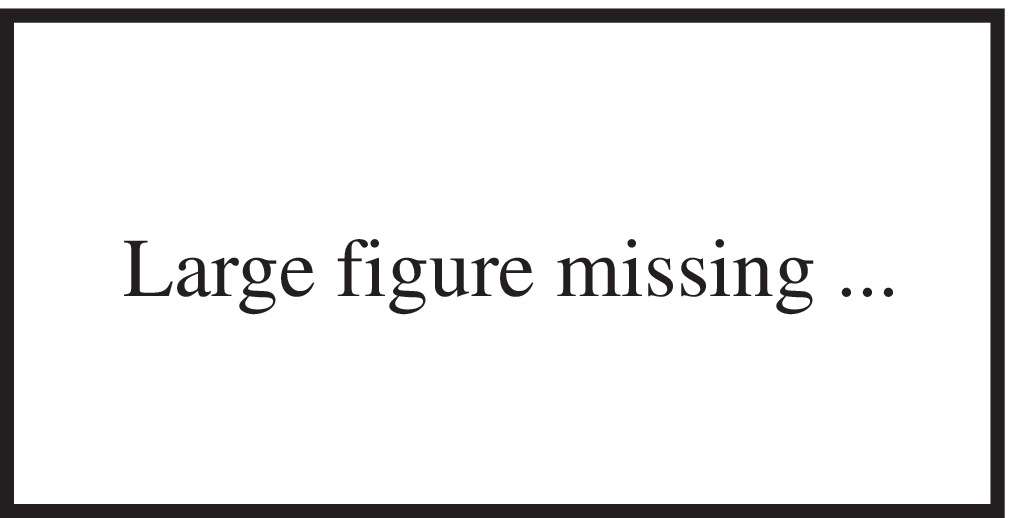}
\end{center}
\caption{\label{fig:TPMS}Visualizations of 10 different cubic
minimal surfaces. Every triply periodic surface divides space into two
labyrinths, which in some cases are represented here by skeletal
graphs.  D and G are the basis for the double-diamond structure and
the gyroid, respectively, which often occur in physical systems.}
\end{figure} 

The best known triply periodic minimal surface (TPMS) is the P-surface
found by Schwarz in 1867 and depicted in \fig{fig:geometries}d.
Schwarz and his students found 5 TPMS, including the cubic cases P,
C(P) and D. Until 1970 no more examples were found, then Schoen
described 13 more \cite{a:scho70}, including the cubic cases G, F-RD,
I-WP, O,C-TO and C(D).  He proved the existence of G by providing an
explicit (Weierstrass) representation, and the existence of the others
was proven in 1989 by Karcher \cite{a:karc89}.  More TPMS have been
discovered by Fischer and Koch \cite{a:fisc87,a:fisc88} and others,
and today many more can be generated by making controlled
modifications to computer models of known TPMS, a method which was
pioneered by Polthier and Karcher
\cite{a:karc96}.  However, these surfaces tend to be rather
complicated, and we will show below that only the simple (as
quantified by the topology index) TPMS are relevant for amphiphilic
systems. In \fig{fig:TPMS} we show visualizations of 10 different TPMS
with cubic symmetry. There representations have been obtained from the
$\Phi^6$-Ginzburg-Landau theory as will be explained below.  For some
of these structures, we also show line-like representations
of the two labyrinths defined by the surface (\emph{skeletal graphs}).
For example, in the case of the Schwarz P-surface, the two skeletal
graphs are the edges of the surrounding cube and the three lines
through the origin.

\begin{table}
\begin{center}
\begin{tabular}{|l|l|l|l|} \hline
     & $\chi$ & $A^*$ & $\Gamma$ \\ \hline
G    & $-8$ & $3 (1 + k_1^2) / 2 k_1 = 3.091444$ & $0.766668$ \\ \hline
D    & $-16$ & $3 / k_1 = 3.837785$ & $0.749844$ \\ \hline
I-WP & $-12$ & $2 \sqrt{3} = 3.464102$ & $0.742515$ \\ \hline
P    & $-4$ & $3 k_1 = 2.345103$ & $0.716346$ \\ \hline
C(P) & $-16$ & $3/k_2 = 3.510478$ & $0.655993$ \\ \hline
\end{tabular}
\end{center}
\caption{Euler characteristic $\chi$, scaled surface area $A^*$
in the conventional unit cell and topology index 
$\Gamma = ({A^*}^3/2 \pi |\chi|)^{1/2}$ 
for those TPMS, for which exact results are known
from Weierstrass representations. Here $k_1 = K(1/2) / K(\sqrt{3}/2)$ 
where $K(k)$ is the complete elliptic integral of the first kind, 
and $k_2 = K(1/\sqrt{3}) / K(\sqrt{2/3})$. Note that often a unit
cell is chosen for D which is the eighth part of the one
chosen here; then one has $\chi = -2$ and $A^* = 1.918893$.}
\label{TableNumbers}
\end{table} 

Different methods can be used to obtain representations of TPMS.
Until recently, the main method were the Weierstrass representation
formulae. For the cubic TPMS, they are known for P, D and G
\cite{a:fodg92b,a:fodg99} as well as for I-WP
\cite{a:lidi90,a:cvij94}.  For each of these TPMS, a fundamental
domain can be identified, so that the rest of the surface follows by
replicating it with the appropriate space group symmetries
($Im\bar{3}m$, $Pn\bar{3}m$, $Ia\bar3d$ and $Im\bar{3}m$,
respectively). The Weierstrass representation is a conformal mapping
of certain complicated regions within the complex plane onto the
fundamental domain:
\begin{equation} \label{WeierstrassDarstellung}
(x_1,x_2,x_3) = Re \int_0^{u + i v} dz\ R(z)\ (1 - {z}^2,i (1 + {z}^2),2 z)
\end{equation}
where $(u,v)$ are the internal (and conformal) coordinates of the
minimal surface. The Weierstrass mapping can be understood as the
inversion of the following composition: first the surface is mapped
onto the unit sphere via its normal, and then the unit sphere is
mapped onto the complex plane by stereographic projection. The
geometrical properties of a surface follow from the Weierstrass
representation as
\begin{equation} \label{WeierstrassDarstellung2}
dA(z) = |R(z)|^2\ (1 + |z|^2)^2\ du dv,\ 
H(z) = 0,\ 
K(z) = \frac{-4}{|R(z)|^2 (1 + |z|^2)^4}
\end{equation}
with $z = u + i v$.  Obviously the (isolated) poles of $R(z)$
correspond to the flat points ($K = 0$) of the minimal surface. Only
few choices of $R(z)$ yield embedded minimal surfaces.  The ones for D
and P have been known since the 19th century from the work of Schwarz:
for D it is $R(z) = (z^8 - 14 z^4 + 1)^{-\frac{1}{2}}$.  P 
follows simply by the Bonnet transformation
$R(z) \to e^{i \theta} R(z)$ with $\theta = 90^\circ$.
\eq{WeierstrassDarstellung2} implies that P and D have the same metric
and the same distribution of Gaussian curvature.  However, since they
map differently into embedding space, they have different space groups
and lattice constants. The gyroid G was discovered in 1970 by Schoen
\cite{a:scho70} as another Bonnet transformation of D, with $\theta =
38.015^\circ$. The Weierstrass representation for I-WP was found only
recently \cite{a:lidi90,a:cvij94}. If one of its poles is chosen to be
at infinity, one has $R(z) = (z (z^4 + 1))^{-\frac{2}{3}}$. In
\tab{TableNumbers} we give exact results for geometrical properties
which have been derived from Weierstrass representations (although for
C(P) no Weierstrass representation is known, these values can be
derived from its complementary relationship to P). Note that the
topology index $\Gamma$ establishes the hierarchy G - D - P within
this Bonnet family.  This sequence corresponds to the connectivity of
the labyrinths defined by the TPMS: G has 3-fold, D 4-fold and P
6-fold coordination, since higher coordinated structures have more
holes.

It is interesting to note that the identification of these structures
in material systems is a difficult enterprise, which depends on the
availability of suitable representations.  Physical representations of
TPMS have been built since the 19th century, using soap films draped
onto wire skeletons (without the wire skeletons surface tension would
shrink these structures into collapse). In 1967 Luzzati and Spegt
noted that certain phases in lipid systems are cubic bicontinuous
\cite{a:luzz67}, but it was only in 1976 that Sciven suggested the
relevance of TPMS as structural models for cubic bicontinuous phases
\cite{a:scri76}.  The Luzzati-Spegt structure was later identified
with the gyroid structure G, and the diamond structure D was
identified both in lipid \cite{a:long83} and diblock copolymer systems
\cite{a:thom86}.  For diblock copolymer systems, the gyroid structure
G was identified in 1994 \cite{a:hadj94}, and little later it was
noted that earlier identifications of cubic bicontinuous phases in
diblock copolymer systems often mistook the gyroid structure G for the
diamond structure D \cite{a:hajd95} (in fact, for diblock copolymer
systems one should rather use the terms \emph{double diamond} and 
\emph{double gyroid}, since in this case two
interfaces are arranged around the corresponding TPMS). Today, the
structures G, D, P and I-WP seem to be non-ambiguously identified in
amphiphilic system, with experimental evidence based mainly on small
angle scattering experiments, electron transmission microscopy, and
swelling and diffusion experiments \cite{a:font90,a:luzz93,a:sedd95}.

Apart from the cases G, D, P and I-WP given above, no more Weierstrass
representations are known for cubic TPMS, so for all other cases
numerical methods have to be used. For example, the \emph{Surface
Evolver} is a software package written by Brakke which allows to
minimize triangulated surfaces for different energy functionals
(compare \sec{sec:www}).  Here we discuss the method of constructing
the isosurfaces of a scalar field $\Phi({\bf r})$, as introduced in
\sec{sec:Landau}.  For our purpose, the usefulness of this model
lies in its rugged energy landscape, which means that many more local
minima exist than the absolute minima corresponding to the lamellar
phase. In physical terms, these additional minima correspond to
modulated phases which are metastable. If started with suitable
initial conditions, the minimization procedure therefore yields
representations which can be used to characterize the structural
properties of these phases.  Therefore this model has been used
repeatedly in order to investigate bicontinuous cubic phases
\cite{a:gomp92a,a:gozd96a,a:gozd96b,uss:schw99}. In particularly, it
has been found that the resulting representations are very close to
TPMS \cite{a:gozd96a,a:gozd96b}.  This finding can be explained as
follows \cite{uss:schw99}: for a triply-periodic cubic structure, the
free energy per unit volume $f = F / V$ follows from the interface
description of \eq{eq:Helfrich} as
\begin{equation} \label{fcurv}
f = \frac{1}{a} \left( \sigma A^* \right) +
\frac{1}{a^3} \left( 2 \kappa \int dA \ H^2
+ 2 \bar \kappa \pi \chi \right)\
\end{equation}
where $a$ is the lattice constant of the conventional unit cell.  Both
terms in brackets are scale invariant, that is they do not depend on
the lattice constant $a$. Since for the Ginzburg-Landau model at hand
the first and second term in brackets is negative and positive,
respectively (compare \sec{sec:Landau}), a balance exists between the
negative surface tension term, which favors small values of $a$, and
the positive curvature contributions, which favors large $a$. If we
make the assumption that the minimization problem can be decomposed in
two independent minimizations, one for lattice constant $a$ and one
for the shape, we can first minimize $f$ for $a$ while considering
$A^*$ and $\int dA \ H^2$ to be constant:
\begin{equation} \label{fmin}
a_{min} = \left( \frac{6 \pi \bar \kappa \chi +
6 \kappa \int dA H^2}{|\sigma| A^*} \right)^{\frac{1}{2}}, \quad
f_{min} = - \left( \frac{4}{27} \right)^{\frac{1}{2}}
\left( \frac{(|\sigma| A^*)^3 }{2 \bar \kappa \pi \chi +
2 \kappa \int dA H^2} \right)^{\frac{1}{2}}\ 
\end{equation}        
Similar mechanisms are always at work in amphiphilic systems and
explain why in contrast to surface tension, bending rigidity
does not necessitate the presence of a scaffold (like a wire
skeleton for soap films) to prevent collapse.  In a second
step, we now minimize $f_{min}$ for shape. If we assume $A^*$ to be
constant, we have reduced our problem to the Willmore problem, and
this is why minimal surfaces with $H = 0$ appear in the framework of
the $\Phi^6$-theory. The final free-energy density can be written as
\begin{equation} \label{fgeom}
f_{min} = - \left( \frac{4}{27} \right)^{\frac{1}{2}}\
\left( \frac{|\sigma|^3}{|\bar \kappa|} \right)^{\frac{1}{2}}\ \Gamma
\end{equation} 
where $\Gamma$ is the topology index defined in \eq{eq:indices}.  We
conclude that the stability of the different TPMS is governed by their
topology indices and that the most favorable bicontinuous cubic phase
is the gyroid G since it has the largest value for $\Gamma$ (compare
\tab{TableNumbers}). Although this conclusion is based on some
assumptions, it is corroborated by a detailed numerical analysis
\cite{uss:schw99}.

In order to obtain representations for a large list of TPMS, we used
the Fourier approach and the theories of black and white space groups
to obtain TPMS as local minima of the $\Phi^6$-theory
\cite{uss:schw99}.  Black and white space groups are also known as
magnetic or Shubnikov space groups and lead to a crystallographic
classification of TPMS \cite{a:fisc87,a:fisc96}. Implementation of
black and white symmetries leads to a considerable reduction in the
degrees of freedoms of the Fourier series. Therefore this method is
computationally cheap and its results are easy to document and to
reuse. A triply periodic surface partitions space into two labyrinths,
which can be considered to be colored black and white. In the
framework of the Ginzburg-Landau theory, black and white correspond to
$\Phi > 0$ and $\Phi < 0$, respectively.  The surface is called
\emph{balanced} if there exists an Euclidean transformation $\alpha$
which maps the white labyrinth onto the black one and vice versa;
otherwise it is called \emph{non-balanced}.  Examples for non-balanced
surfaces are I-WP and F-RD, and the Fourier approach for the
corresponding structure follows the usual rules for the respective
space group \cite{b:shmu96}.  However, if the surface is balanced, the
structure is characterized by \emph{two} space groups: if the colored
structure has space group $\H$, the uncolored structure has space
group $\G = \H \otimes \{ 1, \alpha\}$. $\H$ contains all symmetry
operations of $\G$ that do not interchange the two labyrinths. It is a
subgroup of $\G$ of index 2, that is the quotient group is isomorphic
to the cyclic group of order $2$, $\G / \H \cong {\mathbb Z}_2 \cong \{
1, \alpha \}$.  For example, for the balanced cubic surface defined by
$\cos x + \cos y + \cos z = 0$, the operation interchanging the two
labyrinth is the translation by half a body diagonal, and we have $\H
= Pm\bar3m$ and $\G = Im\bar3m$.

Since $\H$ has index $2$ in $\G$, it follows from the theorem of
Hermann \cite{a:sene90} that $\H$ has either the same point group or
the same Bravais lattice as $\G$.  If $\H$ and $\G$ have the same
point group, their Bravais lattices have to be different.  Thus the
Euclidean operation $\alpha$ has to be a translation which extends one
cubic Bravais-lattice into another.  For the cubic system, there are
only three different Bravais lattices (simple cubic P, body-centered
cubic I and face-centered cubic F), and only two possibilities to
extend one into the other by a translation $\t_{\alpha}$: for
$\t_{\alpha} = a (\x + \y + \z) / 2$ a P-lattice becomes a I-lattice,
and for $\t_{\alpha} = a \x / 2$ a F-lattice becomes a P-lattice. The
condition $\Phi(\r + \t_{\alpha}) = - \Phi(\r)$ leads to reflection
conditions for the reciprocal vectors. For Miller indices $(h,k,l)$,
one finds $h+k+l = 2n+1$ for $P \to I$, and $h,k,l = 2n+1$ for $F \to
P$.  Note that these reflection conditions are similar to the
well-known ones $h + k + l = 2n$ for I and $h+k, h+l, k+l = 2n$ for F.
The P-surface has space group $Pm\bar{3}m$, but a black and white
symmetry with supergroup $Im\bar{3}m$, therefore it is an example for
the case $P \to I$. Thus we have $h+k+l = 2n+1$ and the first Fourier
mode is $(1,0,0)$. Note that the double P structure, which one would
expect for a diblock copolymer system, has no black and white
symmetry, but space group $Im\bar{3}m$, thus one has $h + k + l = 2n$
and the first mode is $(1,1,0)$.
 
If $\H$ and $\G$ have the same Bravais lattice, their point groups
have to be different.  Thus the Euclidean operation $\alpha$ has to be
a point group operation which extends one cubic point group into
another one.  There are five cubic point groups and six ways to extend
one of them into another by some $P_{\alpha}$. The condition
$\Phi(P_{\alpha} \r) = - \Phi(\r)$ leads to more complicated rules
than in the case of identical points groups \cite{uss:schw99}. However,
there are few relevant examples from this class, the most interesting
one being the gyroid G, where $P_{\alpha}$ is the inversion and the
resulting rules are rather simple again (the even part of the Fourier
series vanishes).

\begin{table}
\begin{center}
\begin{tabular}{|l|l|l|l|l|} \hline
& $\H$ & $\G$ & $\H \to \G$ & nodal approximations \\ \hline \hline
G & $I4_132$ & $Ia\bar{3}d$ & $432 \to m\bar3m$ &
$\sin(x) \cos(y) + \sin(y) \cos(z) + \sin(z) \cos(x)$ \\ \hline 
D & $Fd\bar{3}m$ & $Pn\bar{3}m$ & $F \to P$ &
$\cos(x-y) \cos(z) + \sin(x+y) \sin(z)$ \\ \hline
P & $Pm\bar{3}m$ & $Im\bar{3}m$ & $P \to I$ &
$\cos(x) + \cos(y) + \cos(z)$ \\ \hline
C(P) & $Pm\bar{3}m$ & $Im\bar{3}m$ & $P \to I$ &
$\cos(x) + \cos(y) + \cos(z) + 3 \cos(x) \cos(y) \cos(z)$ \\ \hline
\end{tabular}
\end{center}
\caption{Group - subgroup pairs $\G - \H$, the relation $\H \to \G$
between them and nodal approximations for some balanced cubic minimal surfaces.
$\G$ and $\H$ differ either in Bravais lattice or in
point group, but not in both.  Nodal approximations only consider the
space group information given by $\G - \H$. For the simple cases, the first
mode of the corresponding Fourier series is sufficient to obtain
the correct topology.}
\label{nodal1}
\end{table}                                 

\begin{table}
\begin{center}
\begin{tabular}{|l|l|l|} \hline
& $\G$ & nodal approximations \\ \hline \hline
I-WP & $Im\bar{3}m$ & 
$2 [\cos(x) \cos(y) + \cos(y) \cos(z) + \cos(z) \cos(x)]$ \\ 
& & $- [\cos(2 x) + \cos(2 y) + \cos(2 z)]$ \\ \hline
F-RD & $Fm\bar{3}m$ & 
$4 \cos(x) \cos(y) \cos(z)$ \\ 
& & $- [\cos(2 x) \cos(2 y) + \cos(2 y) \cos(2 z) + \cos(2 z) \cos(2 x)]$ \\ \hline
\end{tabular}
\end{center}
\caption{Nodal approximations for some non-balanced cubic minimal surfaces,
for which $\H \equiv \G$. In these cases, more than one mode is needed.}
\label{nodal2}
\end{table}                                 

Fischer and Koch have completely enumerated all 34 cubic
group-subgroup pairs $\G - \H$ with index 2 which are compatible with
cubic balanced TPMS \cite{a:fisc87}. Although there is no way to
completely enumerate all TPMS belonging to a given pair $\G - \H$ in
general, this is possible for a certain subset, that is for all cubic
balanced TPMS which contain straight lines which form a
three-dimensional network \cite{a:fisc87}. This list reads P, C(P), D,
C(D), S and C(Y).  In our work, we implemented Fourier series for
these structures as well as for the gyroid G (which is balanced, but
does not contain any straight lines) and the non-balanced structures
I-WP and F-RD.  In some cases like G, D and P, the first mode of the
Fourier ansatz implementing the correct black and white symmetry is
already sufficient to obtain a representation which is topologically
correct (\emph{nodal approximation}). In all other cases investigated,
the addition of one more mode (with the relative weight fixed by
visual inspection) is sufficient. Nodal approximations for
bicontinuous structures have been discussed first by von Schnering and
Nesper \cite{a:schn91}. In \tab{nodal1} and \tab{nodal2} we give nodal
approximations and the corresponding space group information for some
of the balanced and non-balanced structures investigated.

\begin{figure}
\begin{center}
\includegraphics[width=\textwidth]{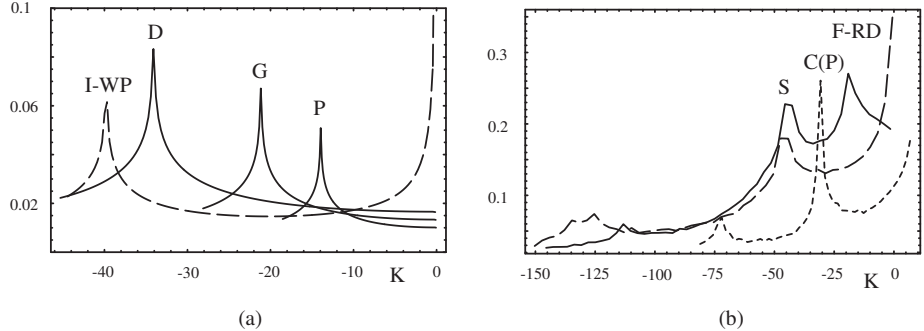}
\end{center}
\caption{\label{fig:kdistributions}Distribution $f(K)$ of Gaussian 
curvature $K$ over the cubic minimal surfaces investigated in the
framework of the $\Phi^6$-model. (a) For the structure G, D, P and
I-WP, the data is obtained from their exact Weierstrass
representations; the agreement with the numerical results from the
$\Phi^6$-model is good (not shown).  (b) For the structures F-RD, S
and C(P), no Weierstrass representation is known, and the data is
obtained from the $\Phi^6$-model.}
\end{figure} 

Implementing the complete Fourier series and numerically minimizing
the $\Phi^6$-functional from \eq{Phi6} with nodal approximations as
initial conditions, we arrived at \emph{improved nodal approximations}
which were tabulated in Ref.~\cite{uss:schw99} with up to six Fourier
modes. The isosurface construction is easily implemented using the
marching cube algorithm. The resulting triangulations can then be used
to investigate geometrical properties of these surfaces.  In
particular, widely used mathematics programs (like Mathematica, Maple
or Matlab) can be easily used to obtain useful representations of TPMS
from (improved) nodal approximations. Our numerical work shows that
the six modes of the improved nodal approximations are sufficient to
decrease the deviation of the curvature properties from the real TPMS
by one order of magnitude compared to the nodal approximations. Also
we measured for the first time the distribution $f(K) = \int dA(u,v)\
\delta(K - K(u,v))$ of Gaussian curvature $K$ over the surfaces. 
For this purpose, we used up to 100 Fourier modes. The
distributions $f(K)$ for different structures are plotted in
\fig{fig:kdistributions}.  In the cases for which Weierstrass
representations are known, the same data has been derived from the
exact representations, and the resulting agreement was very good.  In
general, we found that the surfaces C(D), C(P), F-RD, S and C(Y), for
which no Weierstrass representations are known, are much more
complicated than the cases G, D, P and I-WP. This is evidenced by
larger values of the topology index $\Gamma$, multi-modal
distributions $f(K)$ of Gaussian curvature $K$, and the larger widths
of these distributions.  The latter property can be quantified by
defining a dimensionless variance
\begin{equation} \label{Delta}
\Delta = \frac{\langle ( K - \langle K \rangle )^2 \rangle}
                {\langle K \rangle^2}\ 
\end{equation}
where $\langle \dots \rangle$ means area average.  In
\tab{TableMoments} we give the variance $\Delta$ of the different
distributions. Its value is the same for G, D and P due to the
existence of a Bonnet-transformation between them.  This means that G,
D and P have the narrowest distribution of $K$ (they are most
uniformly curved), and all other structures have a considerably wider
one, with I-WP being the next best structure.

\begin{table}
\begin{center}
\begin{tabular}{|l|l|l|l|l|l|} \hline
& G, D, P & I-WP & S & F-RD & C(P) \\ \hline                       
$\Delta$
& $0.218702$
& $0.482666$
& $0.586079$
& $0.649801$
& $0.842022$ \\ \hline   
\end{tabular}
\end{center}
\caption{Variance $\Delta$ of the distributions of Gaussian curvature $f(K)$.
The values for $\Delta$ are the same for G, D and P due to the
existence of a Bonnet-transformation between them.}
\label{TableMoments}
\end{table}                        

\section{Parallel surfaces}
\label{sec:parallel}

We now consider the case of cubic bicontinuous phases in lipid-water
mixtures. The two examples for the experimental phase behavior of such
systems given in \fig{fig:phasebehavior2} show obvious similarities,
and we will show now that a theoretical description can nicely explain
these \cite{uss:schw00b,uss:schw01a}. It has been shown by a thorough
analysis of electron density maps derived from X-ray data that the
mid-surfaces of the cubic bicontinuous structures in these systems are
very close to minimal surfaces \cite{a:luzz93}. Indeed if one
considers the lipid bilayer as one entity, it has no spontaneous
curvature by symmetry and one expects a minimal surface shape for the
midsurface. Each of the two monolayers of the bilayer has unequal
sides and therefore finite spontaneous curvature.  For lipids
monolayers, spontaneous curvature increases towards the water side as
a linear function of temperature. From the interface point of view,
one expects each monolayer to form a CMC-surface. However, then the
distance to the minimal mid-surface would vary with position and the
amphiphilic tails would have to stretch in order to fill the internal
space of the membrane \cite{a:ande88,a:char90}.  It can be shown in
the framework of a simple microscopic model that the relative
importance of stretching to bending contributions to the free energy
of the bilayer scales as $(a/\delta)^2$, where $a$ is lattice constant
and $\delta$ is tail length \cite{uss:schw01a}. Therefore stretching
is prohibitively expensive and the two monolayers do not form
CMC-surfaces as expected from the curvature energy of
\eq{eq:Helfrich}, but rather parallel surfaces to the minimal
mid-surface.  The free energy of the cubic bicontinuous phases then
follows by specifying \eq{eq:Helfrich} with spontaneous curvature for
two monolayers which are parallel surfaces to a given TPMS. Since the
parallel surface geometry does not allow to completely relax the
bending energy, the overall structure is called \emph{frustrated}
\cite{a:ande88,a:char90,a:dues97}. 

In principle, the analysis in the framework of the parallel surface
model is rather simple, since there exist exact formulae which express
the geometrical properties of the parallel surface as a function of
the geometrical properties of the minimal surface:
\begin{equation} \label{eq:ParallelSurfaces}
dA^{\delta} = dA\ (1 + K {\delta}^2)\ , 
H^{\delta}  = \frac{- K {\delta}}{1 + K {\delta}^2}\ , 
K^{\delta}  = \frac{K}{1 + K {\delta}^2}\ 
\end{equation}                          
where ${\delta}$ is the distance between the two surfaces (that is
amphiphilic chain length). Note that these formulae are an extension
of Steiner's theorem from integral geometry to non-convex bodies,
which is valid as long as the distance ${\delta}$ is smaller than the
smallest radius of curvature of the surface (otherwise the surface
will self-intersect).  Since mean curvature $H$ vanishes on the
reference surface, the bending energy now becomes a function only of
its Gaussian curvature $K$. However, as we have seen above, $K$ is
distributed over the TPMS in a non-trivial way, which makes the
detailed analysis rather complicated. Nevertheless, one central
conclusion can be already made at this point: using
\eq{eq:ParallelSurfaces} in \eq{eq:Helfrich} leads to an effective
curvature energy for the lipid bilayer (to second order in $\delta$) 
\cite{a:port92}
\begin{equation}
  \label{eq:EffectiveBilayerBending} 
F_{bi} = \int dA \left\{ 4 c_0^2
  \delta^2 \kappa + (2 \bar \kappa + 8 c_0 \delta \kappa + 4 c_0^2 \delta^2 \kappa) K
  + 4 \kappa \delta^2 K^2 \right\}\ .
\end{equation}
Thus we see that although effective spontaneous curvature vanishes due
to the bilayer symmetry, the effective saddle-splay modulus,
$\bar{\kappa}_{bi} = 2 \bar \kappa + 8 c_0 \delta \kappa +
O(\delta^2)$, is corrected to higher positive values due to the
presence of the monolayer spontaneous curvature $c_0$. We conclude
that as long as $c_0 \delta \gtrsim - \bar{\kappa} / 4 \kappa$, the
bicontinuous cubic phases are favored over the lamellar phase, since
the preferred curvature of the monolayers translates into a
topological advantage of saddle-type bilayer structures. We also see
that the correction term in $\bar{\kappa}_{bi}$ is linear in $c_0$ and
therefore in temperature $T$. Therefore bicontinuous phases become
more favorable with increasing temperature in general (compare also
\sec{sec:random} on disordered bicontinuous structures).

For a detailed analysis of cubic bicontinuous phases
made from bilayers, we first note that the volume fraction of the
lipid tails (the hydrocarbon volume fraction, which for simplicity we
identify with the lipid volume fraction, since the lipid heads are
rather small) can be calculated as
\begin{equation} \label{hydrocarbonvolume}
v = \frac{1}{a^3} \int_{- \delta}^{\delta} d{\delta'} \int dA^{\delta'}
  = 2 A^* \left( \frac{\delta}{a} \right)
                 + \frac{4 \pi}{3} \chi \left( \frac{\delta}{a} \right)^3
\end{equation}                                       
where again we have used the Gauss-Bonnet theorem.
\eq{hydrocarbonvolume} can be inverted to give $a/\delta$, the lattice
constant $a$ in units of the chain length $\delta$, as a function of
hydrocarbon volume $v$. For small $v$ we find
\begin{equation} \label{LatticeConstant}
\frac{a}{\delta} = \frac{2 A^*}{v} \left( 1 + \frac{1}{12\ \Gamma^2} v^2
+ {\cal O}(v^4) \right)\ .
\end{equation} 
and one can check numerically that $a/\delta = 2 A^* / v$ is an excellent
approximation for $v \lesssim 0.8$. For larger values of $v$, the
surfaces begin to self-intersect and our model becomes
unphysical.  Combining \eq{eq:Helfrich} and \eq{eq:ParallelSurfaces}
yields for the bending energy per unit volume of the two monolayers
(we use a factor $\delta / 4 \kappa c_0^2$ to write this
expression dimensionless and a factor $\delta$ to write $c_0$
dimensionless)
\begin{align} \label{FreeEnergyDensityBicont}
f = v\ \Big\{ \int \frac{dA^*}{A^*} &
\left( 1 -  \Xi(K^*) \left(\frac{v}{\Gamma} \right)^2 \right)^{-1} \\
& \times \left( 1 - \frac{1+c_0}{c_0} \Xi(K^*) 
             \left(\frac{v}{\Gamma} \right)^2 \right)^2
+ \frac{r}{4 c_0^2} \left( \frac{v}{\Gamma} \right)^2 \Big\}\ \nonumber
\end{align}                                                 
where $K^* = K a^2$ is scaled Gaussian curvature, $\delta/a$ has been
replaced by $v / 2 A^*$, $\Xi(K^*)$ has been defined as $K^* A^* / 8
\pi \chi$ (this can be considered to be a local analogue of
$\Gamma^2$), and $r$ as $- \kappa / 2 \bar \kappa$ ($0 \le r \le 1$
due to the restrictions on $\bar \kappa$). The free-energy density $f$
now is a function of lipid volume fraction $v$, spontaneous curvature
$c_0$, the ratio $r$ of the two bending constants, and the
distribution $f(K)$ of Gaussian curvature $K$. The (numerical)
evaluation of this expression is only possible with the knowledge
of the $f(K)$ for all TPMS of interest, which have been
numerically obtained in Ref.~\cite{uss:schw99} as described above.

\begin{figure}
\begin{center}
\includegraphics[width=\textwidth]{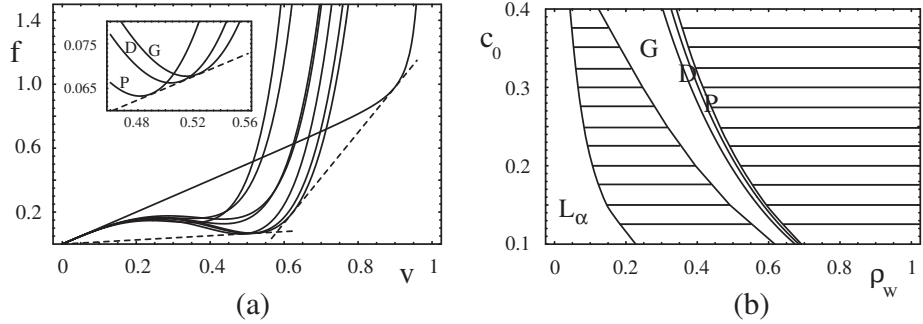}
\end{center}
\caption{\label{fig:langmuir}For cubic bicontinuous phases in 
lipid-water mixtures, the lipid monolayers can be modeled as parallel
surfaces to a minimal mid-surface. (a) Free energy densities $f$ as a
function of lipid volume fraction $v$ for several cubic bicontinuous
phases and the lamellar phase. (b) Theoretical phase diagram as a
function of water volume fraction and (dimensionless) spontaneous
curvature.}
\end{figure} 
 
In the limit of a planar mid-surface (lamellar phase),
\eq{FreeEnergyDensityBicont} simplifies to $f = v$ - the larger the
lipid volume fraction, the more frustrated bending energy per volume
accumulates.  For a full analysis, one also has to include the effect
of thermal fluctuations.  For the lamellar phase, they lead to a
steric repulsion between the interfaces, which lead to an additional
term $\sim v^3 / (1-v)^2$ for the lamellar phase.  For the cubic
bicontinuous phases, steric repulsion is irrelevant since the lateral
restriction on the scale of a lattice constant leads only to small
perpendicular excursions.  However, here the renormalization of the
saddle splay modulus $\bar \kappa$ becomes relevant, and the formula
given in \eq{eq:renorm_kbar} has to be incorporated. The
renormalization of bending rigidity $\kappa$ and spontaneous curvature
$c_0$ is irrelevant here, since thermal fluctuations occur mainly on
the level of the lipid bilayer, for which mean curvature $H$ vanishes.
For the lamellar phase, Gaussian curvature $K$ vanishes as well, and
the renormalization of all material parameters is irrelevant.  Putting
everything together, we can numerically calculate phase diagrams from
the free energy densities of the different phases by using the Maxwell
construction (construction of convex hull).  In \fig{fig:langmuir} we
show both the free energy densities as a function of lipid volume
fraction $v$ (for fixed spontaneous curvature) and the theoretical
phase diagram as a function of water volume fraction and spontaneous
curvature. The main results are in excellent agreement with the
experimental phase diagram shown in \fig{fig:phasebehavior2}: from the
many TPMS considered, only G, D and P are stable, their regions of
stability have the shape of shifted parabolae, and they occur in the
sequence L - G - D - P - emulsification failure. These results can be
understood as follows: G, D and P can achieve the least frustration
since they have the narrowest distribution of Gaussian curvature as
measured by the variance $\Delta$ given in \tab{TableMoments}.  This
prediction has been stated before by Helfrich and Rennschuh
\cite{a:helf90}, but at that time hardly any data was known to support
it. The lipid volume fraction at which they achieve this can be
estimated by setting the dimensionless mean curvature averaged over
the parallel surface
\begin{equation} \label{eq:GeometricalProperty}
\langle H^{\delta} \rangle_{\delta} \delta 
= \frac{\int dA^{\delta}\ H^{\delta} {\delta}}{\int dA^{\delta}}
= \frac{(v / \Gamma)^2}{4 - (v / \Gamma)^2}
\end{equation}   
equal to the spontaneous curvature $c_0$.  Therefore $c_0$ as a
function of $v$ essentially scales as $\sim (1 - \rho_W)^2 /
\Gamma^2$, where $\rho_W = 1 - v$ is the water volume fraction. 
This explains the characteristic shape of the bicontinuous stability
regions in both the theoretical and experimental phase diagrams. Note
that at high temperature, that is large spontaneous curvature,
eventually the hexagonal phase will become stable, which cannot be
treated in the framework presented here. The optimal lipid volume
fraction $v$ follows from \eq{eq:GeometricalProperty} as
\begin{equation} \label{eq:MeanCurvatureAtc0}
v = \left( \frac{4 c_0}{1 + c_0} \right)^{\frac{1}{2}}\ \Gamma\ .
\end{equation}                                               
Therefore the structures G, D and P become stable in the sequence of
their geometry index $\Gamma$, that is as G - D - P. If the water
content corresponds to a average curvature of P which is larger than
the optimal (spontaneous) curvature, some of the water is simply
expelled from the structure, in order to keep the optimal curvature
(emulsification failure). Finally it should be noted that the
Bonnet-transformation connecting G, D and P causes the three
structures to be stable along a triple line (see inset of
\fig{fig:langmuir}). This means that the stability of D and P is very
delicate and can easily be destroyed by additional physical effects.
Therefore we conclude that in the framework of our interfacial
approach, from all TPMS considered the gyroid G is the only phase
which has a robust stability in lipid-water mixtures, since among the
phases with favorable distribution of Gaussian curvature, its
geometrical properties are closest to the ones of the lamellar phase.

\section{Surfaces of constant mean curvature (CMC)}
\label{sec:cmc}

Surfaces of constant mean curvature (CMC-surfaces) have $H = const$
everywhere, so minimal surfaces are a special case of CMC-surfaces. In
contrast to minimal surfaces, CMC-surfaces with finite $H$ can be
compact, but the only CMC-surface which is compact and embedded is the
sphere (for a long time, the sphere was believed to be the only
compact CMC-surface, but in 1986 Wente found the first CMC-torus).
Non-compact embedded CMC-surfaces are cylinder and unduloid, and all
other known CMC-surfaces are doubly or triply periodic.

\begin{figure}
\begin{center}
\includegraphics[width=\textwidth]{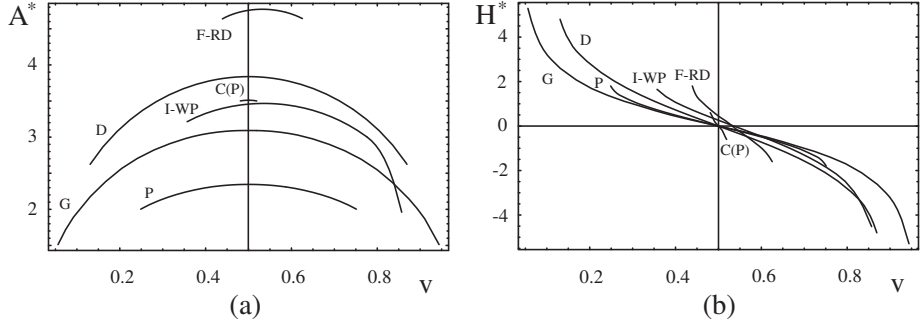}
\end{center}
\caption{\label{fig:cmc}Geometrical data for triply periodic surfaces of 
constant mean curvature: (a) scaled surface area $A^*$ and (b) scaled
mean curvature $H^*$ as a function of volume fraction $v$.  The two
branches are symmetrical for G, D, P and C(P) since their minimal
surface members are balanced.}
\end{figure} 

CMC-surfaces are solutions to the variational problem of minimal
surface area under a volume constraint. This can be shown as follows:
we first introduce a Lagrange parameter for volume, that is pressure
$p$. The corresponding energy is $- p V$. For normal variations
$\delta \phi(u,v)$, we have $\Delta F = 2 \delta \sigma \int dA
\phi(u,v) H(u,v) - p \delta \int dA \phi(u,v) + O(\delta^2)$. If the
surface is required to be stationary in regard to variations in
$\delta$, we obtain the Laplace equation $H = p / 2 \sigma$ and $H$ is
constant over the whole surface. This explains why soap bubbles and
liquid droplets are spheres (respectively spherical caps when bound by
a surface).  In amphiphilic systems, CMC-surface arise in the presence
of spontaneous curvature: like minimal surfaces minimize the Willmore
functional $\int dA H^2$, CMC-surfaces with $H = c_0$ minimize the
functional $\int dA (H - c_0)^2$. In any case, since $H$ is constant
over the surface, surface area and volume now vary in the same way,
and we have $dA / dV = 2 H$.

\begin{table}
\begin{center}
\begin{tabular}{|l|l|l|l|l|l|l|} \hline
              & G & D & I-WP & P & F-RD & C(P) \\ \hline
measured      & 0.2191 & 0.1411 & 0.1385 & 0.2117 & 0.0665 & 0.0466 \\ \hline
approximation & 0.1901 & 0.1465 & 0.1592 & 0.2188 & 0.0906 & 0.1226 \\ \hline
\end{tabular}
\end{center}
\caption{Values for $c = dv(H^*)/dH^*|_{H^* = 0}$, where $v(H^*)$ 
  is the volume fraction of one of the two labyrinths for the
  corresponding family of surfaces of constant mean curvature. The
  measured values follow from our spline interpolation of the
  numerical data of Ref.~\protect\cite{a:ande90}.  In the second row
  we give $- {A_0}^2 / 2 \pi \chi$, a new estimate for $c$ derived in
  the text.}
\label{DataTable}
\end{table}                                                 

All simple TPMS which are of interest for physical reasons are members
of a family of CMC-surfaces \cite{a:karc89}. Each of these families
consists of two branches, corresponding to positive and negative mean
curvatures, which are separated by the minimal surface member. Like
the minimal surface member, each of the the triply periodic
CMC-surfaces of the family partitions space into two intertwined, yet
separate labyrinths, with volume fractions $v$ and $1-v$. Each family
of CMC-surfaces, and therefore the volume fractions of the two
labyrinths and the surface area $A$, is parametrized by $H$. If the
TPMS-member is balanced, then the operation $\alpha$ maps one branch
of the family onto the other, in particular $v(H) = 1-v(-H)$, $v(H=0)
= v_0 = 0.5$ and $A(H) = A(-H)$. For an amphiphilic monolayer of
thickness $\delta$ in a ternary system, $v$ can be identified with the
hydrocarbon volume fraction (that is oil and amphiphilic tails), $1-v$
with the water volume fraction (where we neglect the contributions of
the amphiphilic heads), and $2 A \delta$ with the amphiphile volume
fraction.  In the seminal work by Anderson, Davis, Nitsche and Sciven
from 1990, the CMC-families were numerically constructed for P, D,
I-WP, F-RD and C(P) using finite element methods \cite{a:ande90}.  In
1997, Grosse-Brauckmann numerically constructed the G-family in a
similar way (using the software package \emph{Surface Evolver})
\cite{a:gross97}. For these families, the following data has been
tabulated: volume fraction $v$ as a function of scaled mean curvature
$H^*$ and scaled surface area $A^*$ as a function of $v$.
Rearrangement and interpolation with cubic splines provides smooth
functions $H^*(v)$ and $A^*(v)$. As $v$ is varied away from the value
$v_0$ for the TPMS ($= 0.5$ for balanced structures), one has
\begin{align} \label{TPHSApprox}
H^*(v) & = - \frac{(v - v_0)}{c}       + {\cal O} \left( (v-v_0)^2 \right) \\
A^*(v) & = A_0 - \frac{(v - v_0)^2}{c} + {\cal O} \left( (v-v_0)^3 \right) 
                                  \nonumber
\end{align}         
where the numerical value of $c$ can be extracted from the cubic
splines (compare \tab{DataTable}).  Note that the two relationships
are not independent due to the relation $dA / dV = 2 H$. They are
related through $c$, whose values are well approximated by the
analogous values for the parallel surface companions to the TPMS,
which can be derived as follows \cite{uss:schw00a}: if $\delta$
denotes the perpendicular distance from the minimal to its parallel
surface, to lowest order in $\delta$ the volume fraction $v$ and the
mean curvature $H^*$, averaged over the surface in the unit cell, are
given by $v = v_0 + A^* \delta$ and $H^* = 2 \pi \chi
\delta / A^*$, respectively.  Thus $H^* = 2 \pi \chi (v - v_0) / {A^*}^2$
and $c = - {A^*}^2 / 2 \pi \chi$ for the parallel surface case.  The
corresponding numbers are given in \tab{DataTable}; except for C(P),
the overall agreement with the numerical data for $c$ for the
CMC-surfaces is remarkably good.

\eq{TPHSApprox} is a useful approximation for CMC-surfaces close to
the TPMS, where they behave similar to parallel surfaces.  However, as
mean curvature grows, the numerical data starts to deviate from these
approximations, changes in $v$ and $A^*$ become slower, and a turning
point is reached, where $v$ reaches an extremal value and starts to
decrease again as a function of $H$.  Beyond the turning point, the
surfaces correspond to nearly spherical regions connected by small
necks which resemble pieces of unduloids. Finally these necks
disappear and each branch terminates in an assembly of sphere, which
might be close-packed or self-intersecting.  \fig{fig:cmc} shows the
geometrical data for the families G, D, P, C(P), I-WP, F-RD as
tabulated in the literature \cite{a:ande90,a:gross97}. We do not show
the parts of the data beyond the turning point, as these correspond to
surfaces whose structure is unphysical.

We now consider a ternary mixture of water, oil and amphiphile, where
amphiphiles self-assemble into monolayers with spontaneous curvature
$c_0$. In such a system, all relevant phases (micellar, hexagonal,
lamellar, cubic bicontinuous) can be modeled as CMC-surfaces (spheres,
cylinders, planes, triply periodic CMC-surfaces). This approach has
first been used by Safran and coworkers
\cite{a:safr83,a:safr84,a:wang90}, who in particular discussed the
surfaces from the D-family. In our work \cite{uss:schw00a}, we
extended this analysis to all families of interest, including the
G-family, for which the relevant data has become available only
recently and which features very prominently in experimental systems.
A ternary mixture has two independent degrees of freedom for
concentration, which we choose to be the hydrocarbon volume fraction
$v$ and the ratio $w$ of amphiphile to hydrocarbon volume fraction.
For the formulae given later it is in fact useful to modify the
definition of $w$ and to scale it with the dimensionless spontaneous
curvature $c_0$: $w = \rho_A / v c_0 \delta$. The analysis of \eq{eq:Helfrich}
for CMC-surfaces is simplified considerably by the fact that as mean
curvature $H$ is constant, there is no difference between local and
global curvature properties and the integral over the surface
becomes trivial.  We use a factor $2 \kappa c_0^3$ to rewrite
the free-energy density in the dimensionless form 
\begin{equation} \label{FreeEnergyDensity}
f = \frac{A}{c_0 V} \left( \frac{H}{c_0} - 1 \right)^2
- \frac{2 \pi \chi r}{{c_0}^3 V}\ .
\end{equation}  
For phases with lamellar, cylindrical and spherical aggregates, this
free-energy density can easily be expressed as a function of the
concentration degrees of freedom:
\begin{align} \label{FreeEnergyDensityNonCubic_L}
f_L(w,v) & = w v, \\
\label{FreeEnergyDensityNonCubic_C}
f_C(w,v) & = w v \left(\frac{w}{4} - 1\right)^2, \\
\label{FreeEnergyDensityNonCubic_S}
f_S(w,v,r) & = w v \left[\left(\frac{w}{3} - 1\right)^2
                                    - \frac{r w^2}{9}\right] \ .
\end{align}
Note that only $f_S$ depends on $r$, since the other two structures
have no Gaussian curvature. Since the aggregates are disconnected, the
dependence on $v$ is trivial. The phase boundaries $S-C$, $S-L$, $C-L$
and the emulsification failure are obtained from
Eqs.~(\ref{FreeEnergyDensityNonCubic_L}),
(\ref{FreeEnergyDensityNonCubic_C}) and
(\ref{FreeEnergyDensityNonCubic_S}) to be $w = 24 / (7 - 16 r)$, $w =
6 / (1 - r)$, $w = 8$ and $w = 3/(r - 1)$, respectively.  For example,
for $r = 0$ and increasing $w$, we find the phase sequence L - C - S -
emulsification failure, which is typical for amphiphilic systems (for
simplicity, we identify phase transitions with crossing points of the
free-energy-density curves, and use the Maxwell construction only for
the emulsification failure). In \fig{fig:phasebehavior1} one sees that
experimentally the emulsification failure indeed occurs at constant
$w$ (straight line through water apex). For increasing $r$, the
spherical phase becomes more favorable and finally suppresses the
cylindrical phase.

The structures based on CMC-surfaces discussed here are very different
from the lipid bilayer structures discussed in the preceding section:
now only one monolayer is present, and one of the two labyrinths is
filled with hydrocarbon. In the following, they are called
\emph{single structures}.  It should be noted that for non-balanced
structures, it makes a difference which of the two labyrinth is filled
with hydrocarbon; for example, the I-WP-family generates two single
structures, which we call I and WP.  In ternary amphiphilic systems
there also exists an analogue to the bilayer structures discussed
before, which we call \emph{double structures} and mark with an index
$I$.  Double structures have the same geometries like cubic
bicontinuous phases in diblock copolymer systems. They can be
considered to be TPMS-based bilayer structures where the inner part of
the bilayer has been swollen with suitable solvent. Since each of the
two monolayers has the same spontaneous curvature $c_0$, a double
structure can be modeled as the combination of the two surfaces of a
CMC-family with $H = c_0$ and $H = - c_0$.  Since for not too large
$H$ these two CMC-surfaces essentially correspond to the shrinkage of
one of the two separate labyrinths defined by the minimal surface
member, the two interfaces of a double structure do not intersect.
Note that in principle there exists another class of structures, that
is double structures where oil and water (and therefore amphiphile
orientation and mean curvature) have been reversed. However, since
finite spontaneous curvature selects only phases with curvature
towards one specific side, they are suppressed for energetic
reasons by the lamellar phase.

In our work \cite{uss:schw00a}, we considered 8 different single
structures, which exist for the volume intervals $[0.056,0.944]$ for
G, $[0.131,0.869]$ for D, $[0.249,0.751]$ for P, $[0.481,0.519]$ for
C(P), $[0.357,0.857]$ for I, $[0.143,0.643]$ for WP, $[0.439,0.625]$
for F and $[0.375,0.561]$ for RD. We also considered 6 double
structures, which exist for the volume intervals $[0.112,1.0]$ for
G${}_I$, $[0.262,1.0$] for D${}_I$, $[0.498,1.0]$ for P${}_I$,
$[0.962,1.0]$ for C(P)${}_I$, $[0.624,1.0$] for I-WP${}_I$ and
$[0.818,1.0$] for F-RD${}_I$. Note that the gyroid structures cover
the largest intervals in $v$ for their respective class: there is no
other structure which can incorporate so extreme volume fractions like
the gyroid. Together with the 3 non-cubic phases treated above, we
considered 17 different phases. Since the cubic phases consist of one
connected aggregate, the scaling of the free energy density $f$ with
hydrocarbon volume fraction $v$ will be non-trivial.  For a given
value of $v$, the mean curvature $H(v,a)=H^*(v)/a$ and the surface
area $A(v,a)=A^*(v) a^2$ within a unit cell are determined by the
curves plotted in \fig{fig:cmc}. The amphiphile concentration $\rho_A
= A(v,a)/a^3 = A^*(v)/a$ fixes $a$, so that $a = A^*(v) / \rho_A =
A^*(v) / (w v c_0)$. From \eq{FreeEnergyDensity} we can then
derive
\begin{equation} \label{FreeEnergyDensityCubic}
f_{BC}(w,v,r) = w v \left[\Lambda(v)\ w v - 1\right]^2
+ r\ \frac{(w v)^3}{\Gamma(v)^2}
\end{equation}    
where we have used the definition for the curvature index $\Lambda$
and the topology index $\Gamma$ from \eq{eq:indices}. Note that
now the indices are $v$-dependent. For the 
single structures, $f$ follows by combining using the data
shown in \fig{fig:cmc} with the definitions in \eq{eq:indices}.
For the double structures, the procedure is somehow more
complicated. However, for balanced double structures, 
it becomes simple again, one then can use
\begin{equation} \label{IndicesDouble}
\Lambda_I(v) = \Lambda(v/2)/2, \Gamma_I(v) = 2 \Gamma(v/2),
f_{BC,I}(w,v,r) = 2 f_{BC}(w,\frac{v}{2},r) 
\end{equation}
in \eq{FreeEnergyDensityCubic}.

\begin{figure}
\begin{center}
\includegraphics[width=\textwidth]{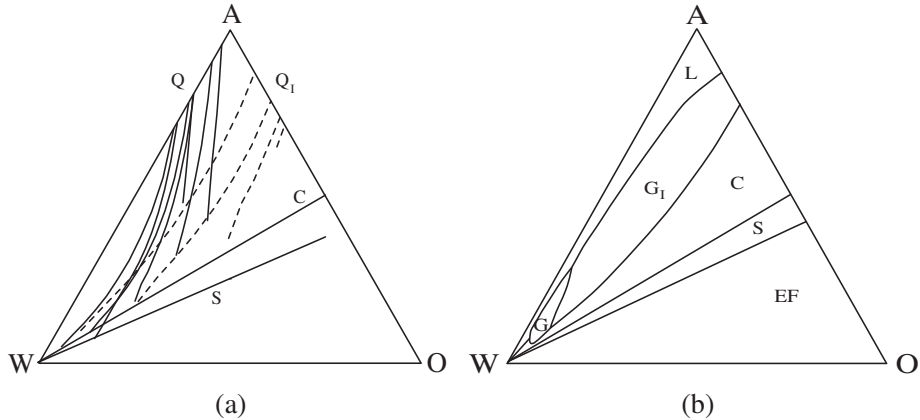}
\end{center}
\caption{\label{fig:gibbstriangle}(a) Lines of vanishing frustration 
for $r = 0$. With increasing amphiphile concentration, the structures
S - C- double cubic - single cubic are stable.  (b) Phase behavior for
$r = 1 / 15$. Now all cubic phases have been suppressed except the two
gyroid phases.  In (a) and (b), $c_0 = 1/6$, which corresponds to
$H_2O/C_{14}/C_{12}E_5$ at $T = 20^{\circ} C$.}
\end{figure} 

For $r = 0$ (vanishing saddle splay modulus $\bar \kappa$), the
analysis of cubic bicontinuous phase behavior becomes rather simple:
in contrast to the case of the parallel surface model discussed in the
last section, now the bending energy given in
\eq{FreeEnergyDensityCubic} can be completely relaxed, namely by
satisfying $w(v) = 1 / v \Lambda(v)$.  This leads to lines of
vanishing frustration in the Gibbs triangle.  For the free-energy
density of the micellar and hexagonal phases
(\eq{FreeEnergyDensityNonCubic_S} and
\eq{FreeEnergyDensityNonCubic_C}, respectively), the lines of
vanishing frustration follow as $w = 3$ and $w = 4$, respectively. All
lines of vanishing frustration are plotted in
\fig{fig:gibbstriangle}a.  Obviously for this case phase behavior is
very complex and degenerated, since every structure considered has
some region of stability around its line of vanishing frustration.
This type of degeneracy caused by the bending energy has been
discussed before \cite{a:Brui92} and leads to the conclusion that
additional physical effects have to be operative, as it is not
observed in experimental systems.

Using \eq{TPHSApprox}, a simple approximation can be derived for the
lines of vanishing frustration of the cubic phases:
\begin{equation} \label{eq:frustration}
w = - c A_0 / v_0 (v - v_0), w_I = - 4 c A_0 / (v - 1)\ .
\end{equation}
Thus the minimal surface case corresponds to the stable solutions for
$w \gg 1$ at $v = v_0$ and $v = 1$, respectively.  For the balanced
single structures and the double structures the hierarchy of the
different phases within the band-like region occupied by a certain
structural type is thus determined by the values of $c A_0$.  Using
the approximation $c \approx - {A^*}^2 / 2 \pi \chi$ derived above, we
find $c A^* \approx \Gamma^2$, where $A^*$ and $\Gamma$ correspond to
the minimal surface members.  Therefore the phase sequence is
approximately determined by the topology index of the minimal-surface
member of each family.  In particular, for a given structural type we
expect to find the sequence G - D - P as a function of either $v$ or
$w$.

For $r > 0$ (negative saddle splay modulus $\bar \kappa$), spheres
become more favorable, cubic phases less favorable and cylinders and
lamellae experience no change in free energy. Therefore the cubic
phases will finally disappear, but our numerical analysis shows that
cubic phases can persist up to $r = 0.2$, in contrast to
earlier work, which predicted $r = 0.1$ \cite{a:wang90}. The reason
for this becomes clear in \fig{fig:gibbstriangle}b where we show the
full phase diagram for $r = 1 / 15 = 0.07$: the only cubic phases
stable here are the two gyroid structures, which have not been
considered before.  There are two main reasons for
their outstanding performance: \eq{FreeEnergyDensityCubic} shows that
their large values for the topology index $\Gamma$ reduces the
energetic penalty caused by $r$, and since they can accommodate extreme
volume fractions, they can compete with other structures at all
relevant concentrations. By comparing the theoretical phase diagram
from \fig{fig:gibbstriangle} with the experimental one from
\fig{fig:phasebehavior2}, we conclude that the experimental system
should correspond to a rather large value of $r$. Earlier work
moreover suggests that incorporation of thermal fluctuations would
favor micellar phases near the water-amphiphilic side and lamellar
phases near the water apex \cite{uss:schw96}. Such a modification is
expected to considerably improve the agreement between the two phase
diagrams.

\section{Random surfaces}
\label{sec:random}

\subsection{Microemulsion and Sponge Phases}

\begin{figure}
\begin{center}
\includegraphics{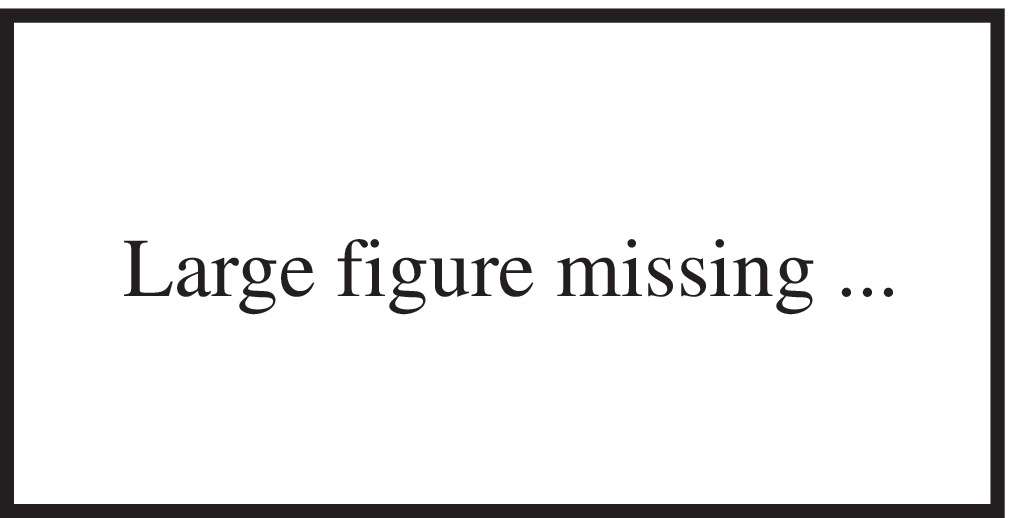}
\end{center}
\caption{\label{fig:freeze_fracture_micro}Freeze-fracture microscopy 
picture of a balanced microemulsion phase. From Ref.~\cite{jahn88}.}
\end{figure}

\begin{figure}
\begin{center}
\includegraphics{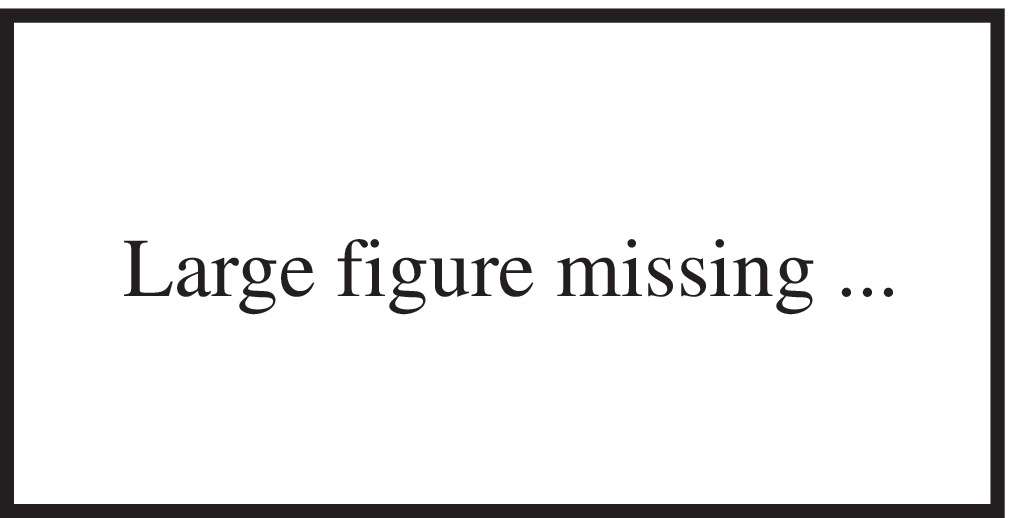}
\end{center}
\caption{\label{fig:freeze_fracture_sponge}Freeze-fracture microscopy 
picture of a sponge phase. From Ref.~\cite{stre92a}.}
\end{figure}

When $\kappa / k_B T$, the bending rigidity in thermal units, becomes
sufficiently low (that is temperature $T$ sufficiently high),
bicontinuous structures can now longer maintain their long-range
crystalline order and melt into a disordered phase, which is
characterized by an exponential decay of correlations in the
interfacial positions.  Such phases have been observed experimentally
for a long time in many binary and ternary amphiphilic systems.  {\em
Microemulsions} are macroscopically homogeneous and optically
isotropic mixtures of oil, water and amphiphiles. On a mesoscopic
scale, they consist of two multiply connected and intertwined networks
of oil- and water-channels, which are separated by an amphiphilic
monolayers. Free-fracture microscopy, where the sample is quickly
frozen, cut, and then studied with an electron microscope, reveals the
intriguing structure of this phase \cite{jahn88}, see
\fig{fig:freeze_fracture_micro}. A similar phase, the {\em sponge
phase}, appears in binary systems of water and amphiphile, where now
the two labyrinths are occupied by water, which are separated by an
amphiphilic bilayer.  The pictures obtained by free-fracture
microscopy \cite{stre92a} are even more suggestive in this case,
because the sample has a preference to break along the bilayer
mid-surface, so that the three-dimensional structure of the membrane
becomes visible. An example is shown in
Fig.~\ref{fig:freeze_fracture_sponge}, which clearly shows the
saddle-like geometry of the amphiphile film.  Therefore, the intuitive
picture of microemulsion and sponge phases as fluid versions of
bicontinuous cubic phases is strongly supported by these experiments.

These phases have been investigated experimentally in considerable
detail over many years. In particular, their phase behavior and
scattering intensities have been studied carefully.  A theoretical
understanding of the statistical mechanics of membranes, however, 
is only beginning to emerge in recent years 
\cite{nels89a,a:Gomp94,peli96,davi96a,gg:gomp97f}. This is no 
surprise, since the statistical mechanics of a surface, which can not
only change its shape, but also its topology in all possible ways, is
extremely complicated. In principle, a partition function of the form
\begin{equation} \label{eq:memb_ensemble}
Z = \sum_{topologies} \int' {\cal D}{\bf R}(\tau) 
    \exp\{{\cal H}[{\bf R}(\tau)]/k_BT\}
\end{equation} 
has to be calculated, where ${\cal D}{\bf R}(\tau)$ denotes an
integration over all possible shapes with parametrization ${\bf
R}(\tau)$ of the surface at fixed topology, where $\tau$ is a
two-dimensional coordinate system on the surface. However, this
integral cannot be just over all possible parametrization ${\bf
R}(\tau)$ of a surface of fixed topology, but has to be restricted to
those parametrizations, which lead to physically different shapes in
the embedding space; this is indicated by the prime. Finally, the
contributions off all different topologies have to be summed over. It
is clear that this problem is sufficiently complex that no exact
solution will be found anytime soon. Therefore, approximations have
to be made in order to get some insight into the behavior of these
phases.

\subsection{Gaussian random fields} 
\label{sec:GRFM}

A very useful approach is to describe the interfaces as isosurfaces of
\emph{Gaussian random fields} (GRF). This corresponds to a Ginzburg-Landau
model as discussed in Sec.~\ref{sec:Landau}, in which the free-energy
functional is taken to be quadratic in the scalar field $\Phi({\bf
r})$. This Gibbs distribution is homogeneous and isotropic, and 
all the functional integrals are Gaussian and can
be performed exactly. In fact, there is a considerable mathematical
literature on the isosurfaces of GRF, see e.g. the book by Adler
\cite{adle81}. In order to comply with the conventions of this
field, we now make two changes to our notation. In the following, area
content is denoted by $S$ (rather than by $A$) and amphiphilic
(hydrocarbon) volume fraction by $\Psi$ (rather than by $v$).

The amphiphilic monolayers in microemulsions have been modelled as
level surfaces of GRF by Berk \cite{berk87,berk91},
Teubner \cite{teub91}, Pieruschka and Marcelja \cite{pier92}, and
Pieruschka and Safran \cite{pier93,pier95a}.  Here the starting point
is a Gaussian free-energy functional of the general form
\begin{equation} \label{eq:GRF}
{\cal H}_0[\phi] = \frac{1}{2} \int d^3q \, 
                 \nu({\bf q})^{-1} \Phi({\bf q}) \Phi(-{\bf q})\ .
\end{equation}
The average geometry of the $\Phi({\bf q})=\alpha$ level surfaces can
be calculated for arbitrary spectral density $\nu({\bf q})$. For the
surface density, $S/V$, the mean curvature $H$, the Gaussian curvature
$K$, and the mean curvature squared $H^2$, the following averages are
obtained \cite{teub91}:
\begin{eqnarray} \label{eq:GRF_SV}
\frac{S}{V} &=& \frac{2}{\pi} \exp\left[-\frac{\alpha^2}{2}\right]
             \sqrt{\frac{1}{3} \langle q^2 \rangle}\ , \\
\label{eq:GRF_K}
\langle K \rangle &=& - \frac{1}{6} \langle q^2 \rangle (1-\alpha^2)\ , \\
\label{eq:GRF_H}
\langle H \rangle &=& \frac{1}{2} \alpha 
                 \sqrt{\frac{\pi}{6} \langle q^2 \rangle}\ , \\
\label{eq:GRF_H2}
\langle H^2 \rangle &=& \langle K \rangle 
              + \frac{1}{5} \frac{\langle q^4 \rangle}{\langle q^2 \rangle}\ 
\end{eqnarray}
where 
\begin{equation}
\langle q^n \rangle = \int \frac{d^3q}{(2\pi)^3} q^n \nu({\bf q})\ .
\end{equation}
The value $\alpha$ of the level cut can be used to describe the
preferred curvature of the membrane, as well as the volume fractions
of oil and water.  Since $\langle H \rangle$ is a linear function of
$\alpha$, compare \eq{eq:GRF_H}, this parameter is proportional to the
spontaneous curvature $c_0$.  In particular, for $\alpha=0$ the mean
curvature of the surface vanishes; this applies to a balanced system,
where $c_0=0$.

For balanced systems, it follows from
Eqs.~(\ref{eq:GRF_SV},\ref{eq:GRF_K}) that the topology index $\Gamma
= \sqrt{8} / \pi = 0.9003$ (compare \eq{eq:indices}). This value is
only slightly larger than the ones for cubic bicontinuous phases
(compare \tab{TableNumbers}), since the balanced random sponge
features only few disconnected parts (note that $\Gamma$ doubles when
the structure is duplicated).  For small curvatures (that is small
$\alpha$), one can use
Eqs.~(\ref{eq:GRF_SV},\ref{eq:GRF_K},\ref{eq:GRF_H}) to derive a
relationship between topology index $\Gamma$ and curvature index
$\Lambda$ which is independent of spectral density $\nu({\bf q})$ and
$\alpha$:
\begin{equation}
\Gamma = \frac{\sqrt{8}}{\pi} 
\left( 1 + \frac{2^8}{\pi^6} \Lambda^4 + \dots \right)\ .
\end{equation}
When the curvature index $\Lambda$ increases since the sponge's
interfaces gain curvature, the topology index increases, too, since
disconnected parts proliferate.

The GRF-approach is most predictive when the Gaussian model of random
interfaces is related to the statistical mechanics of membranes by a
variational approximation \cite{pier93,pier95a}. In this case, the
spectral density $\nu({\bf q})$ in the functional (\ref{eq:GRF}) is
determined by the requirement that the $\phi({\bf r})=0$ level
surfaces mimic the behavior of interfaces controlled by the curvature
Hamiltonian (\ref{eq:Helfrich}) as close as possible. The usual
variational approach employs the Feynman-Bogoljubov inequality,
\begin{equation}
F \le F_0 + \langle {\cal H} - {\cal H}_0 \rangle_0
\end{equation}
where ${\cal H}$ and $F$ are the Hamiltonian and the free energy of
the system of interest, respectively, and ${\cal H}_0$ and $F_0$ the
same quantities of the reference system. In this way, an upper bound
for the true free energy is obtained. This is more complicated in the
case of random surfaces, because the GRF-Hamiltonian is defined
everywhere in space, while the curvature Hamiltonian is only defined
on the level surface. Therefore, the curvature energy does not
restrict fluctuations of the field $\Phi({\bf r})$ away from the level
surface.  In order to suppress such fluctuations, one usually makes
the \emph{mean-spherical approximation}, that is the constraint
$\langle\Phi({\bf r})^2\rangle = 1$ is introduced.

With this variational approach, Pieruschka and Safran \cite{pier93}
have been able to derive the following form for the spectral density
\begin{equation} \label{eq:spectral}
\nu({\bf q}) = \frac{a}{q^4 + bq^2 +c}
\end{equation}
and to relate the parameters $a$, $b$ and $c$ to the curvature elastic
moduli $\kappa$ and $\bar{\kappa}$ and the surface density $S/V$. The
first interesting result is that the spectral density is found to be
{\em independent} of $\bar\kappa$.  Exact expressions for the
parameters can be found in Ref.~\cite{gg:gomp01d}.  To leading order
in $k_BT/\kappa$, the parameters simplify to
\begin{eqnarray} \label{eq:param_a}
a &=& \frac{15\pi^2}{16} \ \frac{k_BT}{\kappa} \ \frac{S}{V}\ ,  \\
\label{eq:param_b}
b &=& \frac{3}{2} \pi^2 \left(\frac{S}{V}\right)^2\ ,  \\
\label{eq:param_c}
c &=& \left( \frac{3\pi^2}{4} \right)^2 \left(\frac{S}{V}\right)^4\ .
\end{eqnarray} 
The spectral density (\ref{eq:spectral}) is equivalent to the 
scattering intensity in bulk contrast. Its Fourier transform yields
the correlation function  
\begin{equation} \label{eq:corr_fun}
\langle \Phi({\bf r}) \Phi({\bf r}') \rangle = 
  \int d^3q \ e^{i{\bf q}\cdot({\bf r}-{\bf r}')} \nu({\bf q}) = 
   \frac{A}{r} \ \exp[-r/\xi] \ \sin(kr)
\end{equation}
where the second equality holds for $|b|<2\sqrt{c}$.
Thus, the correlation function is characterized by two length scales,
the correlation length $\xi$ and the typical domain size $d = 2\pi/k$
of the oil- of water channels, which are obtained from Eqs.~(\ref{eq:spectral})
and (\ref{eq:corr_fun}) to be \cite{teub87} 
\begin{eqnarray} \label{eq:corr_length}
\xi^{-1} &=& \frac{1}{2} \sqrt{2\sqrt{c}+b}\ ,   \\
\label{eq:char_wavevect}
     k   &=& \frac{1}{2} \sqrt{2\sqrt{c}-b}\ .
\end{eqnarray}
With the results (\ref{eq:param_a}), (\ref{eq:param_b}) and (\ref{eq:param_c}),
the asymptotic behavior for small $k_BT/\kappa$ of the dimensionless product 
$k\xi$ is found to be
\begin{equation}
k\xi = \frac{64}{5\sqrt{3}} \ \frac{\kappa}{k_BT} \ .
\end{equation} 
The free energy of the sponge phase can also be calculated from the 
GRF approach. To leading order in an expansion in $k_BT/\kappa$, the
free-energy density $f=F/V$ is found to be \cite{pier95a}
\begin{equation} \label{eq:free_energy_GRF}
f = \frac{\pi^2}{40} [2\kappa-5\bar\kappa] \left(\frac{S}{V}\right)^3 
        - \frac{k_BT}{12} \ln\left(\delta\frac{S}{V}\right) \ .
\end{equation}
This implies that for small membrane volume fractions $\Psi = \delta
S / V$ (where again $\delta$ is the thickness of the amphiphilic
interface), the entropic term dominates over the energy term, and that
the sponge phase becomes unstable in this regime.

We want to end this section with a short discussion of the reliability
of the predictions of the GRF model as a variational approximation for
membrane ensembles. The weak point of this approach is that it is not
clear whether the calculated entropy is actually equivalent to the
physical conformational entropy of the membranes \cite{mors97}. The
main problem is that the curvature energy only controls the shape of
the $\Phi({\bf r})=0$ level surface, while the values of the scalar
field $\Phi$ at all other points in space are not affected by it. The
fluctuations of $\Phi$ in these oil- and water-regions are mainly
determined by the mean-spherical constraint $\langle \Phi^2({\bf r})
\rangle = 1$.  Obviously, an appreciable contribution to the total
entropy arises from the fluctuations of $\Phi$ these `bulk' regions.
This would not affect the predictions of the model as long as the
`bulk' contributions were independent of the interface
positions. Unfortunately, there is no argument so far that this is
indeed the case.

\subsection{Phase behavior of random surfaces} 
\label{sec:phase_random}

From here on we consider only the case of vanishing spontaneous
curvature, $c_0 = 0$, that is balanced microemulsions and sponge
phases. Then the phase behavior is controlled by the bending rigidity
$\kappa$, the saddle-splay modulus $\bar{\kappa}$, the density of
membrane area per volume, $S/V$, and a microscopic cutoff $\delta$,
which can be identified with the thickness of the amphiphilic
interface.  It is very important to realize that for vanishing
spontaneous curvature, phase transition as a function of the
amphiphile volume fraction cannot be understood on the basis of the
curvature energy alone --- i.e. without considering the effect of
thermal fluctuations \cite{a:port92}. The reason is that the curvature
Hamiltonian is conformally invariant in three spatial dimensions,
which implies in particular that it is invariant under a simultaneous
rescaling of all length scales. Since the curvature energy is scale
invariant, the energy {\it density} scales as the third power of an
inverse length, i.e. as $(S/V)^3$. Therefore, the curvature energy of
any given structure --- spherical, cylindrical, lamellar, cubic, or
random --- scales in exactly the same way with decreasing amphiphile
volume fraction, and their relative order is maintained. Therefore,
thermal fluctuations are crucial for these phase transitions.
 
It has been suggested by several authors 
\cite{safr86,cate88a,mors94b,golu94} that the 
free energy of the sponge phase can be obtained by integrating out
the membrane fluctuations on scales less than the typical domain size.
This integration over small-scale fluctuations leads to renormalized,
scale-dependent curvature moduli $\kappa_R(l)$ and $\bar\kappa_R(l)$ 
as given by Eqs.~(\ref{eq:renorm_kappa}) and (\ref{eq:renorm_kbar}), 
respectively, with  
$l/\delta = \Psi^{-1}$, where $\Psi$ is the membrane volume fraction. 
This implies that the curvature Hamiltonian (\ref{Helfrich2}) has to 
be replaced by
\begin{equation} \label{eq:Helfrich_R}
F = \int dA \left\{ \frac{1}{2} \kappa_{+,R}(\Psi^{-1}) (k_1 + k_2)^2 
  + \frac{1}{2}\kappa_{-,R}(\Psi^{-1}) (k_1 - k_2)^2 \right\}
\end{equation}
with 
\begin{eqnarray}
\kappa_{+,R}(\Psi^{-1}) &=& \kappa_+ + \frac{k_B T}{3 \pi} \ln \Psi\ , \\
\kappa_{-,R}(\Psi^{-1}) &=& \kappa_- + \frac{5 k_B T}{12 \pi} \ln \Psi\ . 
\end{eqnarray}
The stability arguments used in \sec{sec:helfrich}
imply that $\kappa_{+,R}$ and $\kappa_{-,R}$ have to be positive 
for the free energy (\ref{eq:Helfrich_R}) to be stable against 
collapse of the structure to molecular scales. Therefore, there
are instabilities at $\kappa_{+,R}(\Psi^{-1})=0$ and 
$\kappa_{-,R}(\Psi^{-1})=0$. 
The latter instability can be identified with the emulsification
failure of the sponge phase, so that the phase boundary is predicted
to occur at \cite{mors94b}
\begin{equation} \label{eq:sponge_stability}
\ln{\Psi} = \frac{6\pi}{5} \ \frac{\bar\kappa}{k_BT} \ .
\end{equation}
This result can be understood intuitively as follows. For sufficiently
large membrane volume fraction, both $\kappa_{+,R}$ and $\kappa_{-,R}$ 
are positive. Therefore, the system tries to minimize $(k_1+k_2)^2$ 
and $(k_1-k_2)^2$. This can be achieved by decreasing both $k_1$ and
$k_2$, i.e. by swelling a given structure as much as possible ---
the lamellar or sponge phase is stable at this value of $\Psi$. 
On the other hand, as soon as $\kappa_{+,R}$ or $\kappa_{-,R}$ become 
negative at some small value of $\Psi$, the free energy can be 
reduced by collapsing the structure. With decreasing length,
however, $\kappa_{+,R}$ and $\kappa_{-,R}$ increase and finally become
positive. Therefore, the collapse stops at a length scale
which is exactly the length scale determined by 
Eq.~(\ref{eq:sponge_stability}).

It is worth mentioning that a similar result follows from the
calculation of the shapes and free energies of passages in lamellar
phases \cite{gg:gomp95c}. Passages are catenoid-like connections
between adjacent lamellae which proliferate close to phase transitions
to disordered bicontinuous phases.  From a detailed calculation, which
takes into account the Gaussian membrane fluctuations, the density,
$\rho$, of passages per unit base area of the stack can be obtained to
be
\begin{equation} \label{eq:passage_density}
\rho = \frac{A_0(\kappa/k_BT)}{\delta^2} \  
        \left(\frac{d}{\delta}\right)^{4/3} \ 
       \exp[4\pi\bar\kappa/k_BT]
\end{equation}
Here, $A_0(\kappa/k_BT)$ is a function which has to be calculated
numerically; it is predicted to increase monotonically with decreasing
$\kappa/k_BT$ in Ref.~\cite{gg:gomp95c}. Therefore, the density of
passages increases with increasing lamellar spacing
$d/\delta=\Psi^{-1}$.  When the average distance between passages
shrinks to the average membrane separation, a transition to the sponge
phase can be expected to occur. This happens at
\begin{equation} \label{eq:passage_limit}
\frac{10}{3} \ln\Psi/\Psi^* = 4\pi \frac{\bar\kappa}{k_BT} 
\end{equation}
where $\Psi^*$ is a function of $\kappa/k_BT$. 
Equations (\ref{eq:sponge_stability}) and (\ref{eq:passage_limit}) 
agree perfectly. 

The consideration of the renormalization of $\kappa$ and $\bar\kappa$
implies that the free-energy density, $f$, of the sponge phase should
behave as \cite{roux90,port91b,roux92a}
\begin{equation} \label{eq:free_energy_renorm}
f = (A + B \ln\Psi) \Psi^3\ .
\end{equation}
As explained above, the overall scaling with $\Psi^3$ derives from the
conformal invariance of the curvature energy.  $A$ represents the
bending energy without thermal fluctuations, and is a linear function
of both $\kappa$ and $\bar\kappa$.  $B$ represents the logarithmic
corrections from the renormalization and is a linear function of
temperature. Both $A$ and $B$ depend on the detailed geometrical
structure of the sponge phase, and thus cannot be obtained from 
simple scaling arguments. It is important to note that the functional
dependence of the free energy (\ref{eq:free_energy_renorm}), which is
based on the renormalization of the curvature elastic moduli due to
{\em small-scale} membrane fluctuations, does not agree with the free
energy (\ref{eq:free_energy_GRF}) of the Gaussian random field model,
which includes the topological entropy of a disordered bicontinuous
phase.

\subsection{Monte Carlo simulations of triangulated surfaces}

In order to go beyond the approximations discussed in Secs.~\ref{sec:GRFM} 
and \ref{sec:phase_random},  
discretized surface models can be investigated by Monte Carlo simulations
\cite{gg:gomp97f,gg:gomp97c,gg:gomp98c,gg:gomp00a}.
Surface triangulations provide the best way of discretizing a surface
as uniformly as possible. The model consists of vertices, which are
connected by bonds in such a way that the bonds form a triangular
network. Several Hamiltonians have been suggested for triangulated
surfaces, such that their shapes and fluctuations are governed in the
continuum limit by the curvature energy from \eq{eq:Helfrich}
\cite{gg:gomp97f}.  Two examples are Gaussian-spring models, in which
neighboring vertices --- the vertices connected by bonds --- interact
with harmonic spring potentials, and tether-and-bead models, in which
the interaction $V(r)$ between neighboring vertices is defined the
potential
\begin{equation}
  V(r) = \left\{ \begin{array}{lll} 
    \infty &             & \ 0 \le r<\sigma_0 \\ 
    0      & \ {\rm for} & \ \sigma_0 < r < \ell_0 \\ 
    \infty &             & \ \ell_0 < r 
   \end{array} \right.
\end{equation}
where $\sigma_0$ is the hard-core diameter of the beads, and $\ell_0$ is 
the tether length. For tether lengths $\ell_0 < \sqrt{3} \sigma_0$, the
network is self-avoiding, because a bead of some distant part of the 
membrane does not fit through the hole between
the three beads of a triangle, even for maximally stretched tethers.

A Monte Carlo simulation of sponge phases requires three types of 
Monte Carlo moves \cite{gg:gomp98c}. The first step is to displace 
selected beads  
by a random vector chosen uniformly in the cube $[-s,s]^3$. Here,
$s$ is determined by the criterion that roughly $50\%$ of trial
moves is accepted according to the Boltzmann weight. The second 
step is to vary the connectivity of the bond network, in order to
allow for diffusion and fluidity of the membrane. This dynamic
triangulation is performed for any two adjacent triangles. The 
bond which forms the common edge of the two triangles is cut, and
a new bond is inserted, which connects the two previously unconnected
beads. The bond can only be cut if every bead retains bonds to
at least three other beads. This procedure guarantees that the network
remains two-dimensionally connected, no holes open up in 
the network, and the topology of the network does not change.  
Finally, the third step is to change the topology of the surface.
This is done by removing two nearby triangles from the surface, and
by connecting the corresponding vertices by a prism of six new 
triangles. Of course, the inverse step is also possible, with a
passage six triangles being removed, and two new triangles inserted
to close the surfaces. The acceptance of both the bond-flip and 
topology-change moves is determined by the Boltzmann weight and is
therefore controlled by the curvature energy. We want to mention
parenthetically that some care has to be taken to find a good
discretization of the bending energy \cite{gg:gomp96b}.

\begin{figure}
\begin{center}
\includegraphics{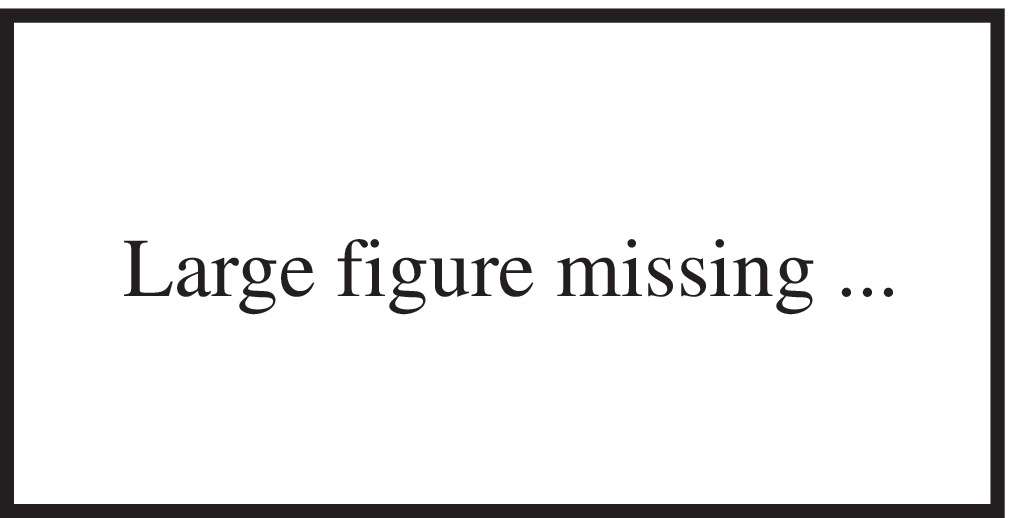}
\end{center}
\caption{\label{fig:sponge_config}A typical membrane configuration 
in a sponge phase for bending rigidity $\kappa/k_BT \simeq 1.6$. The
two sides of the membrane are shaded differently in order to emphasize
the bicontinuous structure of this phase. From
Ref.~\cite{gg:gomp98c}.}
\end{figure}

A typical configuration of a triangulated surface in a cubic box with
parameters $\kappa$, $\bar\kappa$ and $\Psi$ chosen in the stability
region of the microemulsion or sponge phase is shown in
Fig.~\ref{fig:sponge_config}. This configuration nicely demonstrates
the bicontinuous structure of balanced microemulsions and sponge
phases. The saddle-like geometry of the membrane can also be easily
seen. Finally, the figure shows that {\em locally} the structure of
the sponge phase strongly resembles the cubic phases discussed above. 
Therefore, a sponge phase should indeed be considered
as the molten state of the crystalline cubic phase.

A more quantitative comparison with the theoretical approaches of
Secs.~\ref{sec:GRFM} and \ref{sec:phase_random} above can be made by
determining the phase diagram of the randomly-triangulated surface
model, and by calculating the osmotic pressure, $p$, in the
simulations as a function of the membrane volume fraction $\Psi$. From
Eq.~(\ref{eq:free_energy_renorm}), we obtain
\begin{eqnarray} \label{eq:osmotic_press}
p \delta^3/k_BT &\equiv& \frac{1}{k_BT} [\Psi \partial f/\partial \Psi - f] 
              \nonumber \\
  &=& \left[ (2A+B) + 2B \ln\Psi \right] \Psi^3  
\end{eqnarray}
i.e. the same functional dependence as the free-energy density itself.
This dependence of the osmotic pressure is indeed nicely confirmed
by the simulation data \cite{gg:gomp98c}. The simulations therefore 
provide strong evidence for the renormalization of the elastic moduli of the 
curvature model and for the dependence (\ref{eq:free_energy_renorm})
of the free energy on the membrane volume fraction.  

\begin{figure}
\begin{center}
\includegraphics[width=8cm]{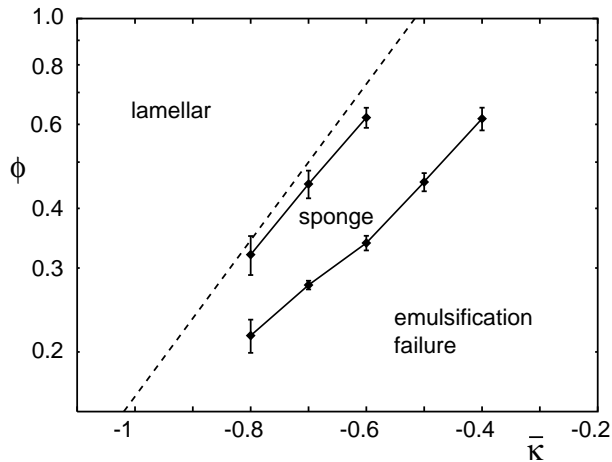}
\end{center}
\caption{\label{fig:phase_diag}The phase diagram as a function of 
membrane volume fraction $\Phi$ and saddle-splay modulus $\bar\kappa$,
for bending rigidity $\kappa/k_BT \simeq 1.6$. Note the logarithmic
scale of the abscissa. The dashed line shows the theoretical
prediction (\ref{eq:sponge_stability}).  From Ref.~\cite{gg:gomp98c}.}
\end{figure}

The phase diagram for fixed bending rigidity $\kappa$ is shown as a 
function of $\bar\kappa$ and $\Psi$ in Fig.~\ref{fig:phase_diag}. 
The simulation
data are compared with the prediction (\ref{eq:sponge_stability}) for 
the phase boundary. Since the slopes of the phase boundaries in this
logarithmic plot agree very well, not only the exponential 
dependence of the membrane volume fraction at the transition on the
saddle splay modulus is confirmed, but also the value of the universal 
prefactor in Eq.~(\ref{eq:sponge_stability}) is strongly supported.

\subsection{Comparison with experiments}

Experimentally, phase diagrams and scattering intensities have
been studied systematically for many different surfactant molecules.
Qualitatively, the agreement with the theoretical approaches
is very reasonable. For example, the scattering curve in bulk
contrast shows a peak at non-zero wave vector in the microemulsion
phase, which moves out and decreases in height with increasing
surfactant concentration. A quantitative comparison, however, is
much more difficult. We want to discuss here three different classes of
experiments, where a such a quantitative comparison has been made.

The first type of experiments are scattering studies in bulk and
film contrast. In bulk contrast, the scattering intensity $S_{ww}({\bf q})$
is proportional to the spectral density $\nu({\bf q})$ of the
Gaussian random field model, compare Eq.~(\ref{eq:spectral}), for
wave vectors $q$ which are not much larger than the
characteristic wave vector $k$ of the domain structure, compare
Eq.~(\ref{eq:char_wavevect}).
The functional dependence of Eq.~(\ref{eq:spectral}) describes the
scattering data in this regime very well, as was first noted by
Teubner and Strey \cite{teub87}, who derived this result on the basis
of a Ginzburg-Landau model, very similar to that introduced in
Sec.~\ref{sec:Landau}. We want
to mention parenthetically, that for wave vectors $k \ll q \ll 1/\delta$,
the intensity is dominated by the scattering from sharp, planar interfaces;
in this limit, the famous Porod law \cite{poro51,teub91} predicts
$I(q) \sim (S/V) q^{-4}$.

In the limit of wave vector $q\to 0$, the scattering intensity in film
contrast is given by \cite{port91b}
\begin{equation}
I(q\to 0) \sim \Psi \left( \frac{\partial p}{\partial \Psi} \right)^{-1}
\end{equation}
where $p$ is the osmotic pressure of Eq.~(\ref{eq:osmotic_press}).
For the free energy (\ref{eq:free_energy_renorm}), this implies
\begin{equation}
[\Psi I(q\to 0)]^{-1} \sim const + \ln\Psi
\end{equation}
Such a behavior has indeed be observed experimentally in Ref.~\cite{port91b}.
However, this result has been questioned by
Daicic et al.~\cite{daic95b,daic95c}. This has lead to a intensive
debate, with arguments against \cite{daic96,daic97} and in
favor \cite{roux96,port97} of the existence of a logarithmic renormalization
of elastic moduli in the sponge phase.

In a second type of experiment, information about the average geometry of the
surfactant film can be extracted from the scattering intensity
in the regime $k < q \ll 1 / \delta$. This information is contained in
the corrections to the asymptotic $1/q^4$ law for smaller values of
the wave vector \cite{teub90}. Experimentally, the average Gaussian
curvature is found to be \cite{chen96a}
\begin{equation}
(V/S)^2 \langle K \rangle = 1.25 \pm 0.10
\end{equation}
which is in excellent agreement with the Gaussian random field result
$(V/S)^2 \langle K \rangle = 1 / \Gamma^2 = -\pi^2/8 = 1.23$
and simulations of the $\Phi^6$-Ginzburg-Landau model
\cite{a:gomp93b,gg:gomp94i}.

The third type of experiment concerns the phase behavior of mixtures
of water and non-ionic surfactant as a function of temperature and
surfactant concentration. As mentioned in \sec{sec:parallel}, the
saddle-splay modulus $\bar\kappa$ is a linear function of temperature
in this case. Therefore, the concentrations at the phase boundaries of
the lamellar and the sponge phase are expected from
Eq.~(\ref{eq:sponge_stability}) to depend exponentially on
temperature. A logarithmic plot of the phase diagram of $C_{12}E_5$ in
water \cite{stre90b}, compare Fig.~\ref{fig:phasebehavior1}, is indeed
consistent with this expectation.

The most detailed information about both the scattering intensities
and phase behavior in these systems has been obtained very recently in
ternary amphiphilic systems of water, oil and non-ionic surfactant
$C_i E_j$, to which small traces of an amphiphilic block copolymer has
been added \cite{jako99,gg:gomp00d,gg:gomp01d,gg:gomp01j,gg:gomp01f}.  
The results obtained in this system provide further, strong evidence for
the existence of a logarithmic renormalization of $\kappa$ and
$\bar\kappa$ in microemulsions and sponge phases.

\section{Summary and outlook}
\label{sec:summary}

In this contribution we discussed the geometrical properties of
surfaces which can be used as structural models for cubic bicontinuous
phases in amphiphilic systems: triply periodic minimal surfaces, their
parallel surfaces and constant mean curvature companions, and
bicontinuous random surfaces.  For each class of surfaces, we showed
how the geometrical properties translate into physical properties of
bicontinuous phases in amphiphilic systems in the framework of an
interfacial description. The surprising success of this approach
relates to the fact that in amphiphilic systems, the solvent has
little physical properties by itself and the free energy is
essentially determined by the interfaces. Although there are several
physical effects which have been neglected in our treatment of
amphiphilic systems, including van der Waals and electrostatic
interactions, it can be concluded that the most essential aspects of
phase behavior are now well understood in terms of the properties of
the underlying geometries.

During the last years, the interest in bicontinuous phases has
increased due to some promising applications in the nano- and
biosciences. For example, amphiphilic self-assembly has been used to
synthesize \emph{mesoporous systems} \cite{n:kres92}, that is porous
material with amorphous walls (usually silica-based) and pore sizes on
the nanometer scale.  As a matter of fact, the combination of
amphiphilic self-assembly and crystallization also seems to be a basic
aspect of biomineralization
\cite{n:mann96}, and there are many algae whose mineral skeletons look
similar to cubic bicontinuous phases \cite{a:hild96}. There is also a
large effort underway to synthesize bicontinuous structures from
graphitic material (\emph{Schwarzites}), which offers the advantage of
atomically smooth walls \cite{n:mack91}. Porous material on the
nanometer scale is not only interesting for its structural properties
(for example use as sieves or catalysts), but also for its electronic
properties (for example use as photonic bandgap material).

In the biosciences, two recent developments involved cubic
bicontinuous phases. It was found that the gyroid phase in the
monoolein-water system provides a functional environment for the
crystallization of integral membrane proteins like bacteriorhodopsin
\cite{a:peba97,a:luec98}. Membrane proteins are notoriously difficult to
crystallize in a three-dimensional array, which however is necessary
for structure determination by X-ray scattering.  And an extensive
analysis of transmission electron micrographs of many biological
specimen has shown that cubic bicontinuous structures are locally
formed in several regions of the cells \cite{a:land95c,a:deng99},
including endoplasmatic reticulum, Golgi apparatus and
mitochondria. These are extended lipid bilayer systems in the cell
which carry a protein machinery responsible for processing and
packaging material inside the cell. In fact it is easy to imagine that
biological cells regulate their lipid composition and therefore their
spontaneous curvature as to achieve geometries favorable for the task
at hand.  In general, cubic bicontinuous phases seem to be ideal space
partitioners if a large amount of active surface area is needed while
simultaneously providing good access for incoming and outgoing
material through the space away from the interface.

What problems are left regarding the subjects of this article ?  From
the viewpoint of mathematics, there are two interesting aspects. On
the one hand, the physical motivation leads to certain variational
problems for which little is known about the corresponding
solutions. One example is the energy functional $\int dA (H - c_0)^2$
with a volume constraint; in principle this problem was subject of
\sec{sec:cmc}, but in lack of adequate mathematical knowledge (this
problem has been treated for vesicles \cite{a:seif96a}, but not for
bicontinuous structures), we used CMC-surfaces as structural models.
On the other hand, some classes of surfaces are well investigated, but
still representations are missing which could be used in physical
modelling. Therefore additional progress in regard to Weierstrass
representations for TPMS, Weierstrass-like representations for their
CMC-companions, and other (numerical) representations would be most
welcome. With a larger repertoire of representations, it should also
become easier to treat non-local effects like van der Waals and
electrostatic interactions; here integral geometry might be helpful in
reducing the dimensions of the corresponding integrals, and
sophisticated summation techniques are needed to account for the
infinite extension of the bulk phases.

From the viewpoint of material sciences, it seems worth to study the
role of boundaries and inclusions.  The term \emph{boundaries}
includes grain boundaries between different, yet adjacent phases, for
example between a hexagonal and a gyroid phase; until now, this
problem has been investigated only for simple geometries.
It also includes the boundary to an external surface
(which might even be curved or chemically structured) and free
boundaries. For example, it has been recently observed experimentally
that the bicontinuous cubic phase in the binary system water -
$C_{12}E_6$ is the only known material system which shows crystal
faces with high Miller indices (\emph{devil's staircase})
\cite{a:pier00}. The term \emph{inclusions} includes several object of
biological and technological relevance, in particular integral
membrane proteins, polymers anchored in the membrane and
(functionalized) nanoparticles. In all of these cases, the framework
described here should provide a good starting point for future
research, which has to bridge the gap between the mathematical
knowledge on the geometrical properties and the physics knowledge of
the material basis of these fascinating structures.

\section{WWW resources}
\label{sec:www}

Here we list some internet addresses which relate to the subject of
this contribution:

\begin{itemize}
  
\item
  \texttt{http://www-sfb288.math.tu-berlin.de/\symbol{126}konrad/}: \\
  Home\-page of Konrad Polthier, one of the pioneers for numerical
  representations and visualizations of minimal surfaces.

\item
  \texttt{http://people.math.uni-bonn.de/kgb/}: \\
  Homepage of Karsten Grosse-Brauckmann, a mathematician working on
  CMC-surfaces, in particular on the G-family.
 
\item \texttt{http://www.susqu.edu/facstaff/b/brakke/evolver/}: \\
The \emph{Surface Evol\-ver} is a public domain software
package written by Ken Brakke for the propagation of triangulated surfaces.
In particular, Ken Brakke shows many examples of TPMS obtained
with the \emph{Surface Evolver}.
 
\item \texttt{http://www.msri.org/publications/sgp/SGP/indexc.html}: \\
  The Scientific Graphics Project offers stunning visualizations of
  minimal and CMC-surfaces, obtained mostly in the framework of
  Weierstrass representations.
 
\item \texttt{http://www.gang.umass.edu/}: \\
  The Center for Geometry,
  Analysis, Numerics and Graphics (GANG), where a lot of work on
  CMC-surfaces takes place.
  
\item
  \texttt{http://www.mpikg-golm.mpg.de/th/people/schwarz/}: \\
  More information on the work described here, including color
  pictures of all structures investigated and (improved) nodal
  approximations.
                  
\end{itemize} 


\end{document}